\pdfoutput=1
\documentclass[11pt,twoside,a4paper,cmspaper,final,collab]{cms-tdr}
\def\svnVersion{c2899d1}\def\svnDate{2025/05/27}\def\cmsCernNoTag{CERN-EP-2025-020}\def\cmsCernDate{\today}\def\cmsMessage{Published in Physical Review D as \href{http://dx.doi.org/10.1103/cm24-665b}{\doi{10.1103/cm24-665b}.}}
\begin{document}\cmsNoteHeader{SMP-22-018}

\cmsNoteHeader{SMP-22-018}

\newlength\cmsFigWidth
\newlength\cmsTabSkip\setlength{\cmsTabSkip}{1ex}
\ifthenelse{\boolean{cms@external}}{\setlength\cmsFigWidth{0.85\columnwidth}}{\setlength\cmsFigWidth{0.4\textwidth}}
\ifthenelse{\boolean{cms@external}}{\providecommand{\cmsTable}[1]{#1}}{\providecommand{\cmsTable}[1]{\resizebox{\textwidth}{!}{#1}}}

\ifthenelse{\boolean{cms@external}}{\providecommand{\cmsLeft}{upper\xspace}}{\providecommand{\cmsLeft}{left\xspace}}
\ifthenelse{\boolean{cms@external}}{\providecommand{\cmsRight}{lower\xspace}}{\providecommand{\cmsRight}{right\xspace}}

\ifthenelse{\boolean{cms@external}}{\providecommand{\cmsFigOneUL}{upper left\xspace}}{\providecommand{\cmsFigOneUL}{left\xspace}}
\ifthenelse{\boolean{cms@external}}{\providecommand{\cmsFigOneUR}{upper right\xspace}}{\providecommand{\cmsFigOneUR}{second from left\xspace}}
\ifthenelse{\boolean{cms@external}}{\providecommand{\cmsFigOneLL}{lower left\xspace}}{\providecommand{\cmsFigOneLL}{third from left\xspace}}
\ifthenelse{\boolean{cms@external}}{\providecommand{\cmsFigOneLR}{lower right\xspace}}{\providecommand{\cmsFigOneLR}{rightmost\xspace}}

\newcommand{\wzg}{\PW\PZ\PGg}
\newcommand{\gbarrel}{\gamma_{\text{barrel}}}
\newcommand{\gendcap}{\gamma_{\text{endcap}}}
\newcommand{\sieie}{\sigma_{\mathrm{i}\eta\mathrm{i}\eta}}
\newcommand{\mll}{m_{\Pell\Pell}}
\newcommand{\mlll}{m_{\Pell\Pell\Pell}}
\newcommand{\mllla}{m_{\Pell\Pell\Pell\PGg}}
\newcommand{\mlla}{m_{\ell_{\PZ1}\ell_{\PZ2}\PGg}}
\newcommand{\ALP}{{\HepParticle{a}{}{}}\Xspace}
\newcommand{\muR}{\ensuremath{\mu_\mathrm{R}}\xspace}
\newcommand{\muF}{\ensuremath{\mu_\mathrm{F}}\xspace}

\cmsNoteHeader{SMP-22-018}
\title{Observation of \texorpdfstring{$\wzg$}{WZgamma} production and constraints on new physics scenarios in proton-proton collisions at \texorpdfstring{$\sqrt{s} = 13\TeV$}{sqrt(s) = 13 TeV}}

\date{\today}

\abstract{
A measurement of the $\wzg$ triboson production cross section is presented. The analysis is based on a data sample of proton-proton collisions at a center-of-mass energy of $\sqrt{s}=13\TeV$ recorded with the CMS detector at the LHC, corresponding to an integrated luminosity of 138\fbinv. The analysis focuses on the final state with three charged leptons, $\Pellpm\PGn\Pellp\Pellm$, where $\Pell = \Pe$ or \PGm, accompanied by an additional photon. The observed (expected) significance of the $\wzg$ signal is 5.4 (3.8) standard deviations. The cross section is measured in a fiducial region, where events with an \Pell originating from a tau lepton decay are excluded, to be $5.48 \pm 1.11\unit{fb}$, which is compatible with the prediction of $3.69 \pm 0.24\unit{fb}$ at next-to-leading order in quantum chromodynamics. Exclusion limits are set on anomalous quartic gauge couplings and on the production cross sections of massive axion-like particles.}

\hypersetup{
pdfauthor={CMS Collaboration},
pdftitle={Measurement of WZgamma production and constraints on new physics scenarios in proton-proton collisions at sqrt(s) = 13 TeV},
pdfsubject={CMS},
pdfkeywords={CMS, standard model, multiboson, new physics}}

\maketitle 

\section{Introduction}\label{sec:intro}

The discovery of the Higgs boson at the CERN LHC~\cite{Aad:2012tfa,Chatrchyan:2012ufa,Chatrchyan:2012ufa_long} experimentally confirmed the principle of electroweak (EW) symmetry breaking~\cite{PhysRevLett.13.321,PhysRevLett.13.508,PhysRevLett.13.585} and the mechanism through which the \PW and \PZ bosons acquire mass.
Following that, many studies have been conducted on physics scenarios beyond the standard model (BSM) as well as the precision measurement of known standard model (SM) processes.
Most of those studies show good agreement with the SM~\cite{CMS:2024phk,ATLAS:2024lda,ATLAS:2024fdw,CMS-SMP-23-004,CERN-EP-2024-093}, motivating BSM searches in previously unexplored event topologies as well as precision measurements of rare SM processes. 
The CMS Collaboration has observed the combined production of three massive gauge bosons at $\sqrt{s}=13\TeV$~\cite{CMS-SMP-19-014}. Recently, also at 13\TeV, $\PW\PGg\PGg$ production has been observed by the ATLAS and CMS Collaborations~\cite{ATLAS-WGG,CMS-SMP-19-013}, and $\PZ\PGg\PGg$ and $\PW\PW\PGg$ production have been observed by CMS~\cite{CMS-SMP-19-013,CMS-SMP-22-006}.
An observation of $\wzg$ production in 13\TeV proton-proton ($\Pp\Pp$) collisions has been reported by the ATLAS Collaboration~\cite{PhysRevLett.132.021802}.

\begin{figure}[ht!]
\centering
      \includegraphics[width=0.20\textwidth]{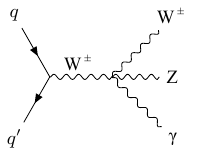}
      \includegraphics[width=0.20\textwidth]{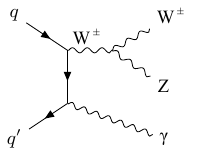}\ifthenelse{\boolean{cms@external}}{\\}{}
      \includegraphics[width=0.20\textwidth]{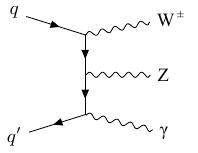}
      \includegraphics[width=0.20\textwidth]{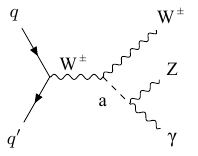}
\caption{Feynman diagrams for $\wzg$ production at LO, including production through a QGC vertex (\cmsFigOneUL), a representative diagram for TGC production (\cmsFigOneUR), and a multiperipheral interaction (\cmsFigOneLL). The \cmsFigOneLR diagram shows the $\wzg$ production including an ALP, denoted $\ALP$, which decays to a \PZ boson and a photon.}\label{fig:wza_feynman}
\end{figure}

Multiboson production processes are sensitive to contributions with triple gauge couplings (TGCs) and quartic gauge couplings (QGCs).
Many BSM models alter these couplings between the gauge bosons.
These contributions can be parametrized through anomalous couplings, which can be described in an effective field theory~\cite{paper_aqgc}.
As one example, the $\wzg$ process is sensitive to potential anomalous TGCs and QGCs (aQGCs).
In addition, the study of $\wzg$ production can be used to search for an axion-like particle (ALP)~\cite{Weinberg:ALP,Wilczek:ALP,ALP1, ALP_atlas,ALP_future,DiLuzio:ALP}.
The axion is a hypothesized pseudoscalar particle, predicted by the Peccei--Quinn~\cite{Axion1,Axion2} solution of the strong charge-conjugation-parity problem in quantum chromodynamics (QCD).
The axion is also a candidate for dark matter. At the LHC, certain types of photophobic ALPs, denoted $\ALP$, can be explored in $\PZ\PGg$ final states~\cite{Craig:2018kne}.
Figure~\ref{fig:wza_feynman} shows representative Feynman diagrams at leading-order (LO) for $\wzg$ production through QGCs, TGCs, multiperipheral interaction, and an ALP, from left to right, respectively.

This paper reports the measurement of the $\wzg$ production fiducial cross section in $\Pp\Pp$ collisions at a center-of-mass energy of 13\TeV corresponding to an integrated luminosity of 138\fbinv, recorded with the CMS detector at the LHC.
Both the \PW and the \PZ bosons are identified via their leptonic decays.
Limits on aQGCs, as well as constraints on the ALP model, are also presented.
Tabulated results are provided in HEPData~\cite{hepdata}.

\section{The CMS detector}\label{sec:cms}

The central feature of the CMS~\cite{Chatrchyan:2008zzk} apparatus is a superconducting solenoid of 6\unit{m} internal diameter, providing a magnetic field of 3.8\unit{T}.
A silicon pixel and strip tracker, a lead tungstate crystal electromagnetic calorimeter (ECAL), and a brass and scintillator hadron calorimeter (HCAL), each composed of a barrel and two endcap sections reside within the solenoid volume. 
Forward calorimeters extend the coverage provided by the barrel and endcap detectors up to pseudorapidities of $\abs{\eta}=5$. Muons are measured in gas-ionization detectors embedded in the steel flux-return yoke outside the solenoid.

Events of interest are selected using a two-level trigger system~\cite{trigger_system}.
The first level (L1), composed of specialized hardware processors, uses information from the calorimeters and muon detectors to select events of interest with a maximum rate of 100\unit{kHz}.
A high-level trigger processor farm decreases this rate to 1\unit{kHz} before storage.
A more detailed description of the CMS detector, together with a definition of the coordinate system and kinematic variables, is reported in Ref.~\cite{Chatrchyan:2008zzk}.

\section{Signal and background simulation}\label{sec:samples}

The $\wzg$ signal is modeled using a sample of inclusive $\Pellpm\PGn\Pellp\Pellm\PGg$ events, where $\Pell = \Pe$ or $\PGm$, simulated with the \MGvATNLO~v2.6.5~\cite{MGatNLO} Monte Carlo (MC) event generator at next-to-LO (NLO) in QCD.
Electrons or muons produced via \PGt decays are included, and the invariant mass of the opposite-charge same-flavor lepton pairs is required to be greater than 4\GeV.
The backgrounds are divided into several categories. (i) The top quark backgrounds, which include $\ttbar\PW$/$\ttbar\PGg$ simulated at NLO with the FxFx jet matching scheme~\cite{Frederix:2012ps}, $\ttbar\PZ$/$\PQt\PZ\Pq$ simulated at NLO using \MGvATNLO~v2.6.5, and $\PQt\PW$ production simulated using the \POWHEG~v2.0~\cite{powheg_tw,POWHEG1,POWHEG2,POWHEG3} generator. (ii) $\PV\PGg$ ($\PV=\PW$ or $\PZ$) backgrounds are simulated at NLO with the FxFx jet matching scheme using \MGvATNLO~v2.6.5. (iii) $\PV\PV$ backgrounds including $\PW\PZ$/$\PZ\PZ$ are simulated at NLO using \POWHEG~v2.0~\cite{powheg_VV}, and the gluon-gluon induced $\PZ\PZ$ production is simulated at LO using the \MCFM~v7.0 generator~\cite{MCFM}. (iv) $\PV\PV\PV$ backgrounds including $\PW\PW\PW$, $\PW\PW\PZ$, $\PW\PZ\PZ$, $\PZ\PZ\PZ$, $\PZ\PZ\PGg$, and $\PZ\PGg\PGg$ are simulated using \MGvATNLO~v2.6.5 at NLO.

The simulation of the aQGC events is performed with \MGvATNLO~2.6.5 at LO. The matrix element reweighting feature in \MGvATNLO provides each event with additional weights corresponding to different theoretical hypotheses, \eg, choice of parameters.
This feature sets the size of the coefficients of any anomalous coupling operators probed in the analysis~\cite{mg5:reweight}.
The simulation of ALP events is performed with \MGvATNLO~2.6.5, using the model from Ref.~\cite{Craig:2018kne}.

The \PYTHIA generator, configured with the CP5 tune, performs parton showering, hadronization, and simulates the underlying event~\cite{CP5}.
The next-to-NLO NNPDF3.1~\cite{NNPDF:2017mvq} parton distribution functions (PDFs) are employed in the simulation, and the CMS detector response is simulated with the \GEANTfour package~\cite{geant4, geant4_2}.
The simulated events include additional $\Pp\Pp$ interactions in the same and neighboring bunch crossings, referred to as pileup (PU).
Simulated events are weighted to match the PU distribution from data, with an average PU of approximately 23, 33, and 32 interactions per bunch crossing during the 2016, 2017, and 2018 data-taking periods, respectively.

\section{Objects reconstruction and event selection}\label{sec:reco}

\subsection{Objects reconstruction}\label{sec:objects}

A particle-flow (PF) algorithm~\cite{paticleflow} is used to reconstruct particles in the event. It combines all subdetector information to reconstruct individual objects and identify them as charged or neutral hadrons, photons, or leptons (PF candidates).
The primary vertex (PV) is taken to be the vertex corresponding to the hardest scattering in the event, evaluated using tracking information alone, as described in Section 9.4.1 of Ref.~\cite{CMS-TDR-15-02}.

Muons are reconstructed using the information in the muon system and the inner tracker within $\abs{\eta}<2.5$ with transverse momentum $\pt>15\GeV$. 
The muon momenta are obtained from the curvature of the corresponding tracks~\cite{Sirunyan_2018}.
Muon candidates are required to satisfy additional identification (ID) criteria, including a minimum number of hits in the muon system and the inner tracker, the quality of the combined fit to a track, the number of matched muon-detector planes, and the consistency between the track and the PV.
The relative isolation is defined as the sum of the charged hadrons and neutral particles in a geometrical cone of $\DR= \sqrt{(\Delta \eta)^2+(\Delta \phi)^2} = 0.4$ centered on the muon trajectory, relative to the muon \pt:
\ifthenelse{\boolean{cms@external}}{
\begin{multline*}
  \text{Iso} = \Bigg(\sum\pt^{\text{charged}} + \\\max\left[0, \sum\pt^{\text{neutral}} + \sum\pt^{\PGg} - \frac{1}{2}\pt^{\text{PU}}\right]\Bigg)/\pt, 
\end{multline*}
}
{
\begin{equation*}
    \text{Iso} = \left(\sum\pt^{\text{charged}} + \max\left[0, \sum\pt^{\text{neutral}} + \sum\pt^{\PGg} - \frac{1}{2}\pt^{\text{PU}}\right]\right)/\pt, 
\end{equation*}}
where $\phi$ is the azimuthal angle, $\sum\pt^{\text{charged}}$ is the scalar \pt sum of charged hadrons originating from the PV, and $\sum\pt^{\text{neutral}}$ and $\sum\pt^{\PGg}$ are the respective scalar \pt sums of neutral hadrons and photons. 
For PU mitigation~\cite{CMS:2019kuk}, $\pt^{\text{PU}}$ is assumed to be half of the scalar \pt sum deposited in the isolation cone by charged particles not associated with the PV.
The factor of 0.5 corresponds approximately to the ratio of neutral to charged hadrons produced in the hadronization of PU interactions.
Three muon selection definitions are used based on the cut-based tight and loose IDs from~\cite{Sirunyan_2018}. The ``tight'' (``veto'') selection is defined by using a high (low) purity working point of the cut-based ID, with the Iso required to be ${<}$0.15\,(0.4) for ``tight'' (``veto'') muons.
The ``loose'' working point is defined by using the high purity working point of the cut-based ID, and the isolation variable is required to be between 0.15 and 0.4.

Electron candidates are reconstructed within $\abs{\eta}<2.5$ with $\pt>15\GeV$.
This involves combining the information from energy clusters deposited in the ECAL and the trajectories fitted in the tracker~\cite{Sirunyan_2021:eg}.
About 35\% of electrons radiate more than 70\% of their initial energy before reaching the ECAL, thus the total energy is evaluated by adding back all bremsstrahlung photons compatible with originating from the electron track, which is implemented through ``super-clustering'' algorithms~\cite{Sirunyan_2021:eg}.
To reduce contamination, \eg, from electrons originating in jets, the electron candidates are required to fulfill additional ID criteria based on the following variables: the relative amount of energy deposited in the HCAL; a match of the trajectory in the inner tracker with that in the supercluster of the ECAL; the number of missing hits in the inner tracker; the consistency between the track and the PV; and $\sieie$~\cite{Sirunyan_2021:eg}, a shower shape variable that quantifies the spread in $\eta$ of the electromagnetic shower in the ECAL.
Electron candidates identified as originating from photon conversion are rejected.
Three working points are defined according to the ID efficiency~\cite{Sirunyan_2021:eg}.
A ``tight'' working point is used to select electrons in the final state, defined by the CMS cut-based medium ID~\cite{Sirunyan_2021:eg}.
A much less restrictive working point is used to perform the additional electron veto, referred to as ``veto'', defined by the cut-based veto ID.
A third working point with an efficiency located between tight and veto, referred to as ``loose'',  defined by the cut-based loose ID, is used to select electron candidates from data to mimic the jet-induced electron background.
The relative isolation variable for electrons is defined similarly as for muons, with the exception that the contribution from PU effects is estimated using the effective area approach~\cite{Sirunyan_2021:eg}. The isolation requirement applied to electrons is included in the cut-based ID.

Photons are identified as ECAL energy clusters not linked to the extrapolation of any charged-particle trajectory to the ECAL~\cite{Sirunyan_2021:eg}.
They are required to have $\pt>20\GeV$ and be within $\abs{\eta}<2.5$, while the transition region between the barrel and endcap ECALs of $1.444<\abs{\eta}<1.566$ is excluded, because the reconstruction of a photon object in this region is not optimal.
Photons located in the barrel region, $\abs{\eta}<1.444$, and the endcap region, $1.566<\abs{\eta}<2.5$, will be referred to as $\gbarrel$ and $\gendcap$, respectively.
An isolation requirement using a similar definition as described above is applied with $\Delta R=0.3$.
The cut-based medium ID~\cite{Sirunyan_2021:eg} is used to select ``tight'', promptly produced, photons.
The loose photon criteria is defined by inverting the $\sieie$ and the isolation requirements for the tight selection. 
The PU suppression is performed in a similar way as for electrons.
An additional veto is applied on electrons reconstructed as photons.

Jets are reconstructed using PF objects and the anti-\kt jet clustering algorithm~\cite{antikt} with a distance parameter of 0.4.
The energies of charged hadrons are determined from a combination of their momenta measured in the tracker and the matching of ECAL and HCAL energy deposits, corrected for the response of the calorimeters to hadronic showers.
The energy of neutral hadrons is obtained from the corresponding corrected ECAL and HCAL energies.
To reduce the contamination from PU, charged PF candidates with $\abs{\eta}<2.4$ are excluded from jet clustering when they are associated with PU vertices~\cite{CMS:2019kuk}, in a process called ``charged hadron subtraction''~\cite{paticleflow}.
The contribution from neutral PU particles to the jet energy is corrected based on the projected area of the jet on the front face of the calorimeter.
Additional ID criteria on the energy fractions of charged and neutral components, as well as a veto on overlaps with leptons ($\DR>0.4$), are applied to jet candidates.
A standard ``tight'' working point is adopted in the analysis.
Corrections are applied to the jet energy as a function of jet $\eta$ and $\pt$ to account for detector response nonlinearities, contributions from PU, and residual differences between the jet energy scale in data and simulation~\cite{jer}.
The \PQb~quark jet ID uses the tight working point~\cite{antikt} of the deep machine learning version of the combined secondary vertex \PQb~tagging algorithm DeepCSV~\cite{Sirunyan:2017ezt}, which combines the information from displaced tracks with the information from secondary vertices associated with the jet and on the possible presence of a soft muon in the event from the semileptonic decay of the \PQb~quark.

The missing transverse momentum vector \ptvecmiss is computed as the negative weighted vector~\pt sum of all the PF candidates in an event, and its magnitude is denoted as \ptmiss~\cite{CMS:2019ctu}.
The \ptvecmiss is modified to account for corrections to the energy scale of the reconstructed jets in the event. A ``tag-and-probe'' procedure~\cite{tagandprobe} is used to measure factors that correct for data-to-simulation differences in the trigger, reconstruction, and selection efficiencies.

\subsection{Event selection}\label{sec:selection}
Collision events are collected using a combination of triggers~\cite{CMS-TRG-19-001} that require the presence of one or two electrons or muons. 
The $\pt$ threshold of the single-lepton trigger is 27\,(24)\GeV for electrons (muons); for the double-lepton triggers with same flavor leptons, the $\pt$ thresholds of the leading and subleading leptons are 23\,(17) and 12\,(8)\GeV.
For the double-lepton triggers with different-flavor leptons, the $\pt$ threshold of the leading lepton is 23\GeV, regardless of the flavor, whereas the $\pt$ threshold of the subleading lepton is 12 or 8\GeV depending on the data-taking period.
The combination of these triggers leads to a nearly 100\% trigger efficiency in the phase space considered by this study.
In 2016 and 2017, partial mistiming of signals in the forward region of the ECAL endcap detectors ($2.75 < \abs{\eta} < 3$) resulted in L1 triggers being wrongly associated with the previous bunch crossing~\cite{cms:l1_trigger}.
Since rules for L1 triggers forbid two consecutive bunch crossings to trigger the detector, events with mistimed signals can self veto, which can lead to a decrease in the L1 trigger efficiency.
This effect is not included in the simulation, instead a correction is implemented using a factor that represents the probability
of the event not having mistimed signals.

A selected event in the $\wzg$ signal region (SR) is required to have exactly three isolated leptons, following the tight working points introduced in Section~\ref{sec:objects}.
The leptons can be electrons or muons, and they can be direct \PW or \PZ boson decay products, or come from an intermediate \PGt lepton decay.
The leading and subleading leptons in \pt used to reconstruct the \PZ boson candidate are referred to as $\Pell_{\PZ1}$ and $\Pell_{\PZ2}$, respectively.
The remaining lepton, associated with the \PW boson, is referred to as $\Pell_{\PW}$.
The $\pt$ requirements on the leptons are: $\pt^{\Pell_{\PZ1}}>25\GeV$, $\pt^{\Pell_{\PZ2}}>15\GeV$, and $\pt^{\Pell_{\PW}}>25\GeV$.
A \PZ boson candidate with invariant mass satisfying $\abs{m_{\Pell_{\PZ1}\Pell_{\PZ2}} - m_\PZ}<15\GeV$ is required to be reconstructed from two opposite-sign charge and same-flavor (OSSF) leptons among the three tight leptons.
If there is more than one combination of OSSF leptons satisfying the \PZ boson candidate requirement, the one with invariant mass closest to the \PZ boson mass is chosen.
The highest $\pt$ photon satisfying the tight selection with $\pt^\PGg>20\GeV$ is chosen as the signal photon.
The invariant mass of the three leptons, $\mlll$, is required to be larger than 100\GeV to reduce the contribution from $\PZ\PGg$ production in which the photon converts and produces an additional electron.
The invariant mass of any same-flavor lepton pair, $\mll$, is required to be larger than 4\GeV to avoid low-mass resonances.
The invariant mass of the photon and the lepton associated with the \PW boson, $m_{\PGg\Pell_{\PW}}$, is required to be outside the interval (75, 105)\GeV when $\Pell_{\PW}$ is an electron, to reduce the contribution from Z boson events with electrons.
Due to the presence of neutrinos in the \PW boson decay, the selected events are required to have $\ptmiss>30\GeV$.
Events with a fourth lepton satisfying the veto working point are rejected to reduce the $\PV\PV$ and $\PV\PV\PV$ backgrounds.
To reduce contributions from top quark processes, events with a \PQb-tagged jet with $\pt^{\PQb}>30\GeV$ and $\abs{\eta^{\PQb}}<2.4$ are vetoed.
The region used for aQGC constraints is the same as the SR, with the exception that $\pt^{\PGg}>60\GeV$.

Control regions (CRs) are defined to constrain or to validate the background models. A CR enriched in events with nonprompt leptons is obtained by inverting the $\PQb$ jet veto and \PZ boson mass window, $\abs{m_{\Pell_{\PZ1}\Pell_{\PZ2}} - m_\PZ}>15\GeV$.
The photon requirement is also removed and the remaining selections are the same as for the SR.
Similarly, a second CR enriched in events with nonprompt photons is defined by requiring only two tight OSSF leptons with a veto on the presence of a third loose lepton.
Also, the SR $\ptmiss$ requirement, the $\PQb$~jet veto, and \PZ boson mass window selection are inverted, and at least one tight photon with $\pt>20\GeV$ is required.

The $\PZ\PZ$ CR is defined by inverting the fourth lepton veto in the SR definition.
A fourth lepton with $\pt>15\GeV$ satisfying the tight working point is required, and the requirements on the photon are removed.
The $\ptmiss$ requirement is also inverted.
Two pairs of OSSF leptons are required, each having $\abs{m_{\Pell\Pell} - m_\PZ}<15\GeV$.
The normalization of the simulated samples used to model the $\PZ\PZ$ background is constrained using this CR.

The selection on $m_{\PGg\Pell_{\PW}}$ is not required in the three CRs.
A summary of all the selection criteria for the various regions is shown in Table~\ref{tab:selections}.

\begin{table*}[ht]
    \centering
    \topcaption{Summary of the event selections in the SR, nonprompt CRs, and the $\PZ\PZ$ CR. The nonprompt CRs are used to validate and constrain the nonprompt lepton and photon contributions, and the $\PZ\PZ$ CR is used to constrain the $\PZ\PZ$ contribution. A ``\NA'' indicates that no requirement is placed on the corresponding observable. The aQGC SR is the same as the SR with the exception that $\pt^{\PGg}>60$\GeV.}
    \cmsTable{
    \begin{scotch}{ccccccccccc}
        Region & $N_\ell$ & $N_\gamma$ & $N_{\rm OSSF}$ & $N_{b tag}$ & \ptmiss & $\pt\{\ell_{Z1},\ell_{Z2},\ell_W,\ell_4\}$ & min$(m(\ell\ell`))$ & $\abs{m(\ell_{Z1}, \ell_{Z2})-m_\PZ}$ & $m(\ell_{Z1},\ell_{Z2},\ell_W)$ & $m(\ell_W,\gamma)$ \\
        & & & & & [\GeVns{}] & [\GeVns{}] & [\GeVns{}] & [\GeVns{}] & [\GeVns{}] & [\GeVns{}] \\
        \hline
        SR & =3  &  $\geq$ 1  & $\geq$ 1  & =0  & $>$30  & $>$\{25, 15, 25, -\} & $>$4 & $<$15 & $>$100 & $<$75 or $>$105 \\
        ZZ CR & =4  & \NA & $\geq$ 1  & =0 & $<$30  & $>$\{25, 15, 25, 15\} & $>$4 & $<$15 & $>$100 & \NA \\
        Nonprompt $\Pell$ CR & =3  & \NA & $\geq$ 1  & $>$0  & $>$30  & $>$\{25, 15, 25, -\} & $>$4 & $>$15 & $>$100 & \NA \\
        Nonprompt $\PGg$ CR& =2 &  $\geq$ 1 & =1 & $>$0 & $<$30  & $>$\{25, 15, -, -\} & $>$4 & $>$15 & \NA & \NA \\
    \end{scotch}%
    }\label{tab:selections}
\end{table*}

\section{Background estimation}\label{sec:bkg}
The dominant backgrounds to $\wzg$ production in the SR stem from nonprompt photon, nonprompt lepton, and $\PZ\PZ$ production.
The former two contributions arise from misidentified objects that are not direct products of the hard scattering, and are usually produced during the subsequent parton shower and hadronization.
The probabilities of nonprompt objects passing the ID requirements are typically low, however, the contributions can still be significant because of the large rate of heavy-quark jets at the LHC.
These nonprompt objects are typically not modeled well in simulation. Consequently, we estimate their contributions directly from data.
The contribution of $\PZ\PZ$ production comes from two parts. In one, a $\PZ$ boson decays into an electron-positron pair and either one of them is misidentified as a photon, denoted by $\Pe\to\PGg$.
Alternatively, one of the four final-state leptons is not within the acceptance or does not pass the ID requirements, while the photon is produced from final-state radiation.
The $\PZ\PZ$ contribution is estimated using simulation, as introduced in Section~\ref{sec:samples}, which will be constrained in the SR through a simultaneous fit with the $\PZ\PZ$ CR.

The background from nonprompt photons arises mainly from events with a jet misidentified as a photon, \eg, $\PW\PZ$+jets or $\PQt\PZ$+jets.
The method of Refs.~\cite{ZGjj_fullrun2,WG_fullrun2} used to estimate this background includes measuring and applying a per-event extrapolation factor in data, which varies as a function of $\pt^{\PGg}$.
The measurement of the extrapolation factor is performed by extracting the fraction of jets misidentified as photons in a region with two or more tight leptons, of which two form a $\PZ$ boson candidate.
The fraction of nonprompt photons is determined via a template fit to the photon $\sieie$ distribution, where values are typically small for prompt photons and large for nonprompt photons.
The prompt template is taken from simulation, and the nonprompt template from a sideband region of the photon isolation variable.
The resulting factors are used to weight events in the application regions: events selected with the same lepton requirements as the nominal SR or nonprompt CR, but with the photon candidate required to pass the loose selection and fail the tight.

The background from nonprompt leptons is mainly comprised of $\PQt\PQt\PZ$ and $\PQt\PQt\PW$ production as well as other jet-enriched processes with jets misidentified as leptons.
The estimation of nonprompt leptons is performed in a similar fashion to that for the nonprompt photons, with the exception that nonprompt leptons produced via $\PGt$ decays are estimated using simulation.
The extrapolation factor is defined as: $\mathrm{W}_{\alpha} =f_{\Pell} / (1-f_{\Pell})$, where $f_{\Pell}$ is the lepton misidentification rate, defined as the ratio of the number of events with leptons passing the tight working point to the total number of events with leptons passing the tight or loose working points.
The $f_{\ell}$ is measured as a function of the $\eta$ and $\pt$ of the leptons and is performed in a dijet-enriched region. The events are required to have only one electron or muon passing the tight or loose working point, at least one tight jet, and small $\ptmiss$.
In addition, a combination of single-lepton triggers with a minimal $\pt$ of 8\GeV is required.
The contribution from prompt leptons, \ie, leptons from $\PW$+jets and $\PZ$+jets, is subtracted from both the numerator and denominator using simulation when calculating $f_{\Pell}$.
To estimate the nonprompt lepton background, these factors are applied as weights to events in three application regions, defined by those events that satisfy the full analysis selection, with the exception that one, two, or three of the leptons pass the loose working point but fail the tight one.
The event weight is calculated by multiplying the factors corresponding to each lepton in events that pass the loose selection but fail the tight selection.
The final contribution of nonprompt leptons is obtained by summing contributions from the three categories. The possible overlap between the three application regions is removed.

Other backgrounds, including top quark and multiboson production, are estimated from MC simulation as mentioned in Section~\ref{sec:samples} and are normalized using their most precise theoretical cross sections. 
The possible double-counting between the different background categories is removed.

\section{Systematic uncertainties}\label{sec:sys}
Systematic uncertainties that affect the measurements arise from experimental sources, such as detector effects and the methods used to compute higher-level quantities, \eg, efficiencies, and from variations in theoretical inputs, such as the choice of the renormalization (\muR) and factorization (\muF) scales and the choice of the PDFs.
Each systematic uncertainty is quantified by evaluating its effect on the normalization and shape of relevant kinematic variables in the signal and background categories, and is propagated to the final distributions on a bin-by-bin basis.
The log-normal distribution, recommended~\cite{CAT-23-001} for implementing multiplicative corrections, is used to model the dependence of the event yields on the systematic uncertainties.

The systematic uncertainties in the measurement of the integrated luminosity are 1.2\%, 2.3\%, and 2.5\% for the 2016, 2017, and 2018 data-taking periods, respectively~\cite{lumi2016,lumi2017,lumi2018}.
Accounting for correlations, the overall uncertainty for the 2016--2018 period is 1.6\%.
The effect of mismodeling the PU is estimated by varying the total inelastic cross section used in the reweighting procedure by $\pm4.6\%$~\cite{CMS:2019kuk}.
This uncertainty is treated as fully correlated among different data-taking years.
The uncertainties in the correction factors accounting for the ECAL mistiming in 2016 and 2017 MC events vary by 1--4\%, and are treated as correlated across different processes and bins but uncorrelated between the 2016 and 2017 data sets.
The systematic uncertainties in the trigger efficiency are measured to be ${\approx}$1\% and are correlated among the three data-taking years, and the uncertainty related to the limited samples for this measurement is smaller than 0.4\% and is uncorrelated among the three years.
The uncertainties in the lepton and photon ID and isolation efficiency scale factors are ${\approx}$2\% and are correlated among different data-taking years.
The systematic uncertainty in the electron (photon) momentum correction is ${\approx}$2\,(3)\% and is correlated across different processes and data-taking years, whereas the uncertainty in the muon momentum correction is negligible.

The uncertainties in the jet energy scale (JES) and jet energy resolution (JER) are calculated from simulated events by rescaling and spreading the jet \pt~\cite{jer}, and propagating the bin-by-bin effects in the variables.
The uncertainties in the event yield due to the JES and JER amount to 1--4\%.
The JES uncertainty components are considered to be either fully correlated or fully uncorrelated across the different data-taking years, depending on the underlying source, while the JER is treated as uncorrelated across the different years. Both JER and JES uncertainties are treated as correlated across processes.

An overall systematic uncertainty in the nonprompt photon background estimate is defined as the sum in quadrature of the systematic uncertainties from several distinct sources.
An uncertainty arising from the choice of the isolation variable sideband is evaluated by estimating the nonprompt photon fraction with a set of alternative choices of the isolation sideband.
A nonclosure uncertainty is defined by performing the template fits using simulated events and comparing the results with the known fractions.
The uncertainty arising from the choice of the prompt template is assessed by comparing nonprompt fractions obtained from two different choices of prompt template, specifically $WZ\PGg$ or $Z\PGg$.
The nonclosure uncertainty in the endcap region is larger than in the barrel region and increases as the photon \pt increases.
The overall systematic uncertainty in the nonprompt photon background is 12--22\%, dominated by the nonclosure effect.
Similarly, the dominant uncertainty in the nonprompt lepton estimate is associated with the nonclosure, which is calculated by comparing two yields, one from the $\PGg$+jets events and the other from the $\PGg$+jets events where the nonprompt lepton rates are applied to events with a lepton that passes the loose, but fails the tight working point.
The other requirements of the main event selection are applied.
The uncertainty associated with the nonprompt lepton background is 30\% for both electrons and muons and is correlated across the data-taking years.
Additionally, the statistical uncertainty from the extrapolation factor calculation is included.

The uncertainties related to the limited number of simulated events, or to the limited number of
events in data CRs, affect the signal and background predictions.
They are uncorrelated across different processes, and across bins of a single distribution. 

Theoretical uncertainties arise from the choice of \muR and \muF, as well as the choice of PDFs.
The uncertainty due to missing higher orders in QCD is estimated through independent changes in the \muR and \muF scales up and down by a factor of two relative to their nominal value in each event, under the condition that $1/2 \leq \muR/\muF \leq 2$. 
Those changes are propagated to the weights of the MC events.
The maximal difference with respect to the nominal yield obtained using the central event weights is taken as the measure of uncertainty.
The uncertainties in the PDFs are estimated by combining the expectations from all of the contributions in the NNPDF3.1 set of PDFs, according to the procedure described in Ref.~\cite{Butterworth:2015oua}.
For the signal, the scale uncertainty is 5--9\% and the PDF uncertainty is 3--5\%, depending on the value of $m_{\PZ\PGg}$.

All the above systematic uncertainties are applied to both the measured signal rate relative to the SM prediction (signal strength) and to the search for BSM physics.
All systematic uncertainties, except those arising from the trigger and lepton ID efficiencies, are assumed to be correlated between the electron and muon channels.
A summary of the contributions to the total uncertainty in the signal strength is given in Table~\ref{tab:nuisance}.

\begin{table}[htb]
  \centering
  \topcaption{Summary of the relative contributions of related groups of uncertainties to the value of the signal strength in the measurement of the SM $\wzg$ signal.}
  \begin{scotch}{lc}
  Uncertainty source    &  $\Delta \mu/\mu_{\text{obs.}}$ \\
  \hline
  Statistical             &   17\%    \\
  Theory                  &   7.6\%    \\
  Nonprompt-background estimation    &   6.6\%    \\
  Photon related          &   5.5\%    \\
  Electron related        &   4.1\%    \\
  Integrated luminosity   &   1.7\%   \\
  Pileup                  &   $<$1\%   \\
  Muon related            &   $<$1\%   \\
  \end{scotch}\label{tab:nuisance}
  \end{table}
  
\section{Results and interpretations}\label{sec:results}
This section presents the measured signal strength, its corresponding significance, and the production cross section of the $\wzg$ process. Limits on the aQGCs are also determined.
Finally, the results are interpreted in the context of a photophobic ALP production model. These results have been determined using the CMS statistical analysis tool \textsc{Combine}~\cite{CAT-23-001}. In order to avoid bias, all of the observed results and any sensitive distributions were blinded until all of the analysis methods were frozen.

\subsection{Measurement of \texorpdfstring{$\wzg$}{WZgamma} production}\label{sec:smwzg}

\begin{figure*}[htbp]
\centering
      \includegraphics[width=0.49\textwidth]{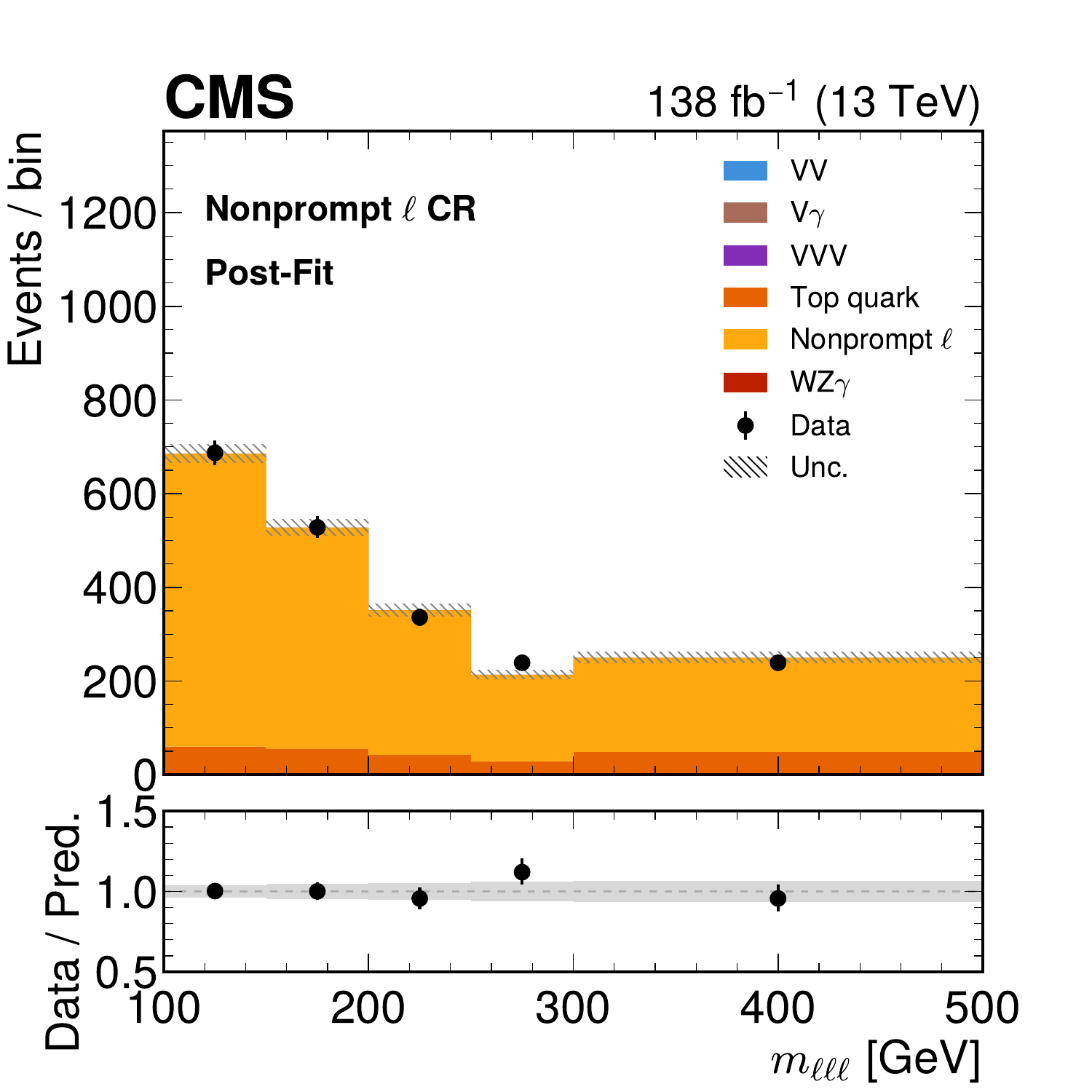}
      \includegraphics[width=0.49\textwidth]{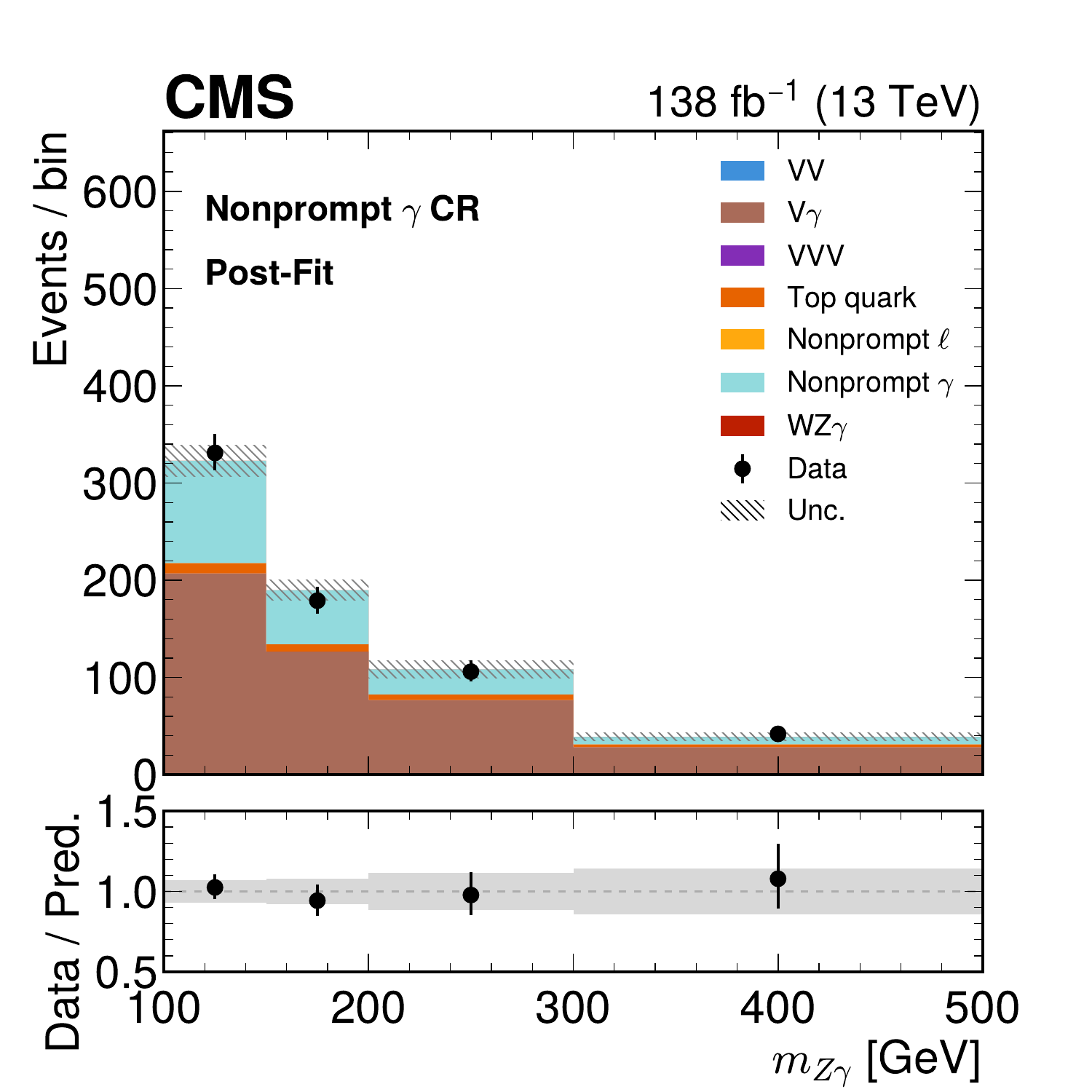}
      \includegraphics[width=0.49\textwidth]{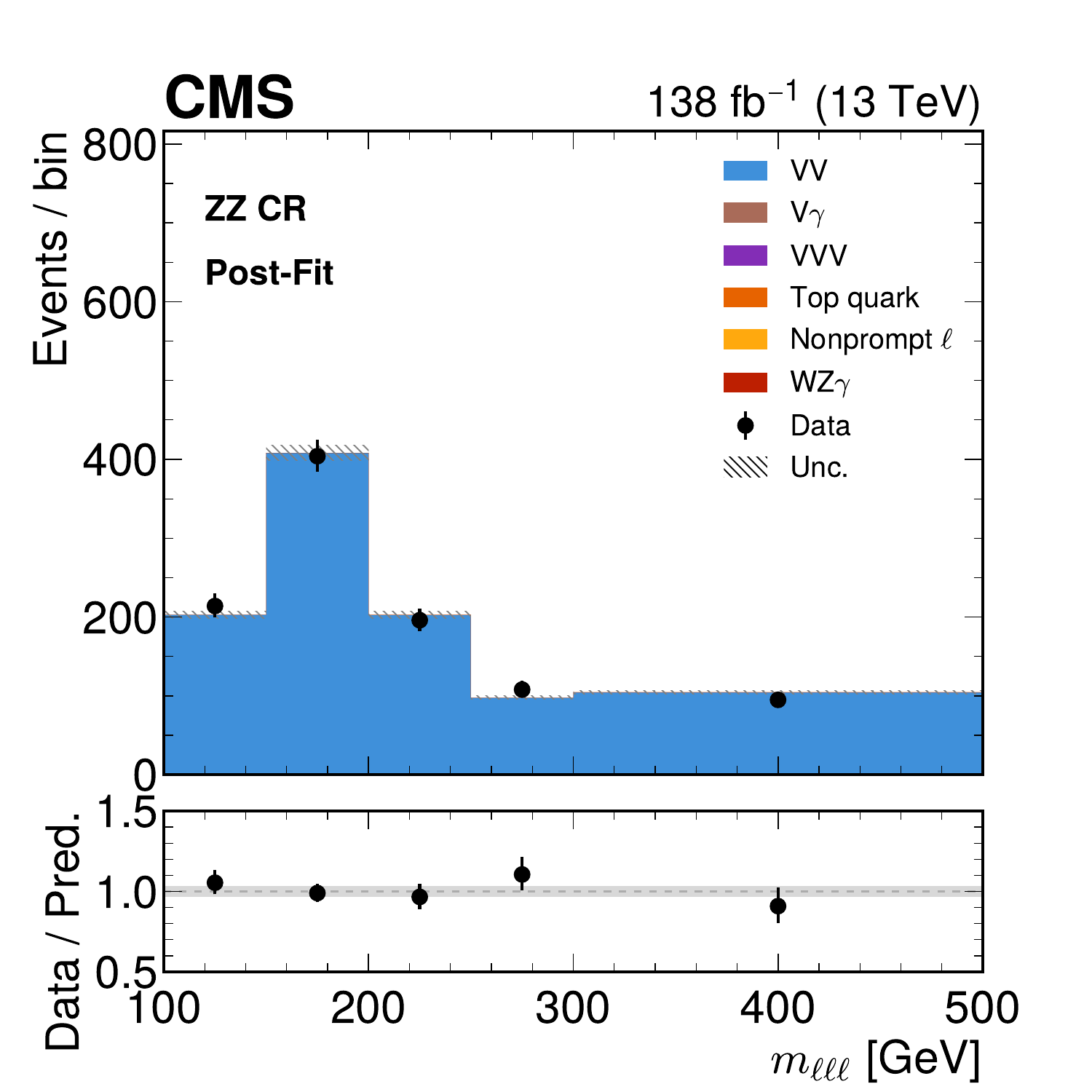}
      \includegraphics[width=0.49\textwidth]{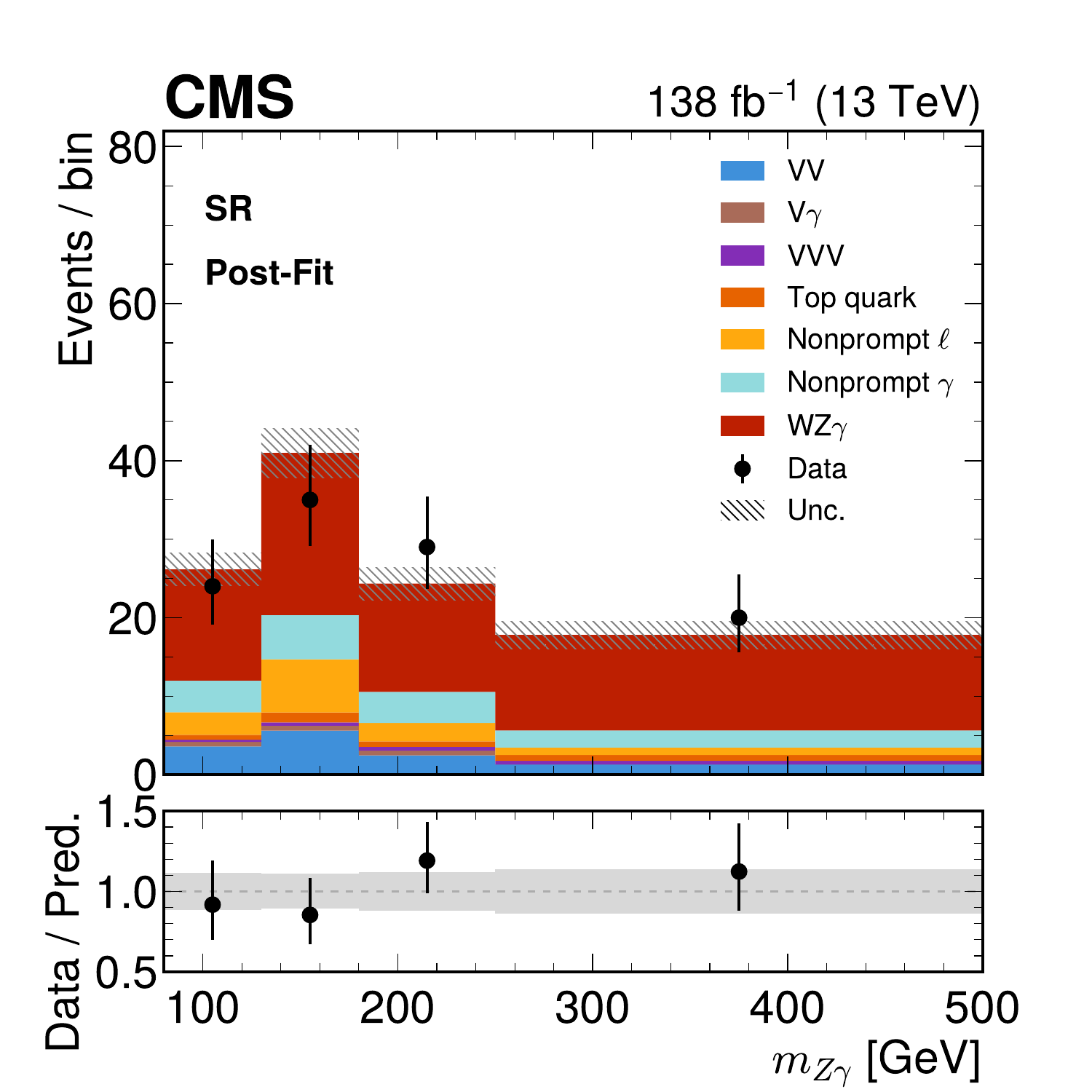}
      \caption{The distributions of the variables used in the simultaneous fit for the nonprompt \Pell CR (upper left), nonprompt \PGg CR (upper right), $\PZ\PZ$ CR (lower left), and SR (lower right) after the fit to the data. The black points with error bars represent the data and their statistical uncertainties, whereas the shaded band represents the total uncertainties. The bottom panel in each figure shows the ratio of the number of events observed in data to that of the total SM prediction. The last bin of each plot has been extended to include the overflow contribution.}\label{fig:prefit}
\end{figure*}

The SM $\wzg$ signal extraction is performed using a simultaneous binned likelihood fit in one-dimensional distributions in $m_{Z\PGg}$ (SR and nonprompt photon CR) and $m_{\Pell\Pell\Pell}$ (nonprompt lepton CR and ZZ CR).
The CRs are included in the fits to constrain the dominant backgrounds.
The data in the SR and all CRs, together with the background and signal yields after the fit to data, are presented in Table~\ref{tab:yields}.
The distributions after the combined fit in each region are shown in Fig.~\ref{fig:prefit}.

\begin{table*}[htbp]
    \centering
    \topcaption{The number of events in data and predictions after the combined fit for the relevant processes in the SR and CRs. All analysis uncertainties are included.
}
    \newcolumntype{x}{D{,}{\,\pm\,}{3.2}}
    \begin{scotch}{cxxxx}
    Process & \multicolumn{1}{c}{SR} & \multicolumn{1}{c}{Nonprompt $\Pell$ CR}& \multicolumn{1}{c}{Nonprompt $\PGg$ CR} & \multicolumn{1}{c}{ZZ CR} \\
    \hline
    $\wzg$ signal  & 60.8,3.5 & 0.66,0.09 & 0.20,0.04 & 0.017,0.008 \\
    $\PV\PV$    & 12.99,0.32 & 1.86,0.12 & 0.164,0.017 & 1016,12 \\
    $\PV\PGg$    & 1.8,0.9 & 4.7,2.1 & 438,27 & 0.011,0.006 \\
    $\PV\PV\PV$   & 1.6,0.4 & 0.36,0.11 & 0.011,0.007 & 0.10,0.04 \\
    Top   & 3.3,0.6 & 227,15 & 27.0,5.9 & 0.30,0.04 \\
    Nonprompt $\Pell$ & 12.9,2.8 & 1792,34 & \multicolumn{1}{c}{$<$0.01} &  \multicolumn{1}{c}{$<$0.01}  \\
    Nonprompt $\PGg$ & 15.8,2.2 &  \multicolumn{1}{c}{$<$0.01} & 195,19 &  \multicolumn{1}{c}{$<$0.01} \\
    [\cmsTabSkip]
    Total prediction & 109,5 & 2027,33 & 660,21 & 1016,12\\[\cmsTabSkip]
    Observed & \multicolumn{1}{c}{108} & \multicolumn{1}{c}{2029} & \multicolumn{1}{c}{658} & \multicolumn{1}{c}{1017} \\
    \end{scotch}\label{tab:yields}
\end{table*}

The $\wzg$ signal normalization is allowed to vary without an external constraint in the fit.
The normalization of $\PZ\PZ$ (gluon-gluon and quark-antiquark 
 induced $\PZ\PZ$ are considered together) is allowed to vary freely since it is well constrained from the data in the CR.
The nonprompt photon and nonprompt lepton backgrounds are derived directly from the observed events, and hence they are fixed to their yields and constrained within the related uncertainties.
Systematic uncertainties are included as nuisance parameters in the likelihood fit.
All of the fitted nuisance parameters are consistent with their nominal values.
All the other background processes are fixed to their SM cross sections, subject to the uncertainties described in Section~\ref{sec:sys}.
The measured signal strength is $\mu^{\text{SR}}_{\text{obs.}}=1.47 \pm 0.15\syst \pm 0.11\thy\pm 0.25\stat=1.47\pm0.31$. 
Considering the background-only hypothesis, this corresponds to an observed (expected) significance of 5.4 (3.8) standard deviations.

\subsection{Fiducial cross section measurement}\label{sec:total_xs}
The leptonic $\wzg$ production cross section measurement is performed in a fiducial phase space region based on particle-level quantities, which is designed to emulate the selection performed at the reconstruction level.
The lepton selected at the generator level is ``dressed'' by adding the four-momenta of any photons with $\DR(\Pell,\PGg)<0.1$ to the four-momentum of the lepton.
Three dressed leptons (electrons or muons), with $\pt>15\GeV$ and $\abs{\eta^{\Pell}}<2.5$ are required, with at least one OSSF pair to reconstruct a \PZ boson candidate.
In cases where multiple OSSF lepton combinations are possible, the pair with the invariant mass closest to the $\PZ$ boson mass
 is selected.
The mass of the \PZ boson candidate is required to be within the interval of 60 to 120\GeV.
The photon candidate is required to have $\pt > 15\GeV$, $\abs{\eta^\PGg}<2.5$, and a separation from each lepton of $\DR(\Pell,\PGg)>$ 0.3.
A photon isolation selection is also applied, which requires the total scalar \pt sum of all stable particles within a cone size $\DR=0.4$ around the photon direction to be less than 50\% of the photon \pt.
Events where an electron or muon originates from a tau lepton decay are excluded.
The fiducial cross section is extracted using the same fitting strategy as the fit used to determine the signal significance, with the following exception: events that pass the SR selection but fail the fiducial selection are regarded as a background. 
The predicted SM cross section for this $\wzg$ fiducial region, obtained using the \MGvATNLO~2.6.5 event generator in NLO, is $\sigma_{\text{SM}}=3.69\pm 0.15\,(\text{PDF}) \pm 0.19\,(\text{scale})\unit{fb}$.
The measured fiducial cross section is $\sigma_{\text{obs.}}= 5.48 \pm 0.57\syst \pm 0.10\thy \pm 0.95\stat = 5.48 \pm 1.11\unit{fb}$, where the theoretical uncertainties include those from the renormalization and factorization scale variations, PDFs, and parton shower modeling in all simulated samples. The measured value is compatible with the SM prediction.

\subsection{Limits on anomalous quartic gauge couplings}\label{sec:aqgc}
The effects of BSM physics can be parametrized in a generic way through a collection of linearly independent higher-dimensional operators in an effective field theory~\cite{paper_aqgc}. 
Reference~\cite{Eboli:2006wa} proposes nine independent charge-conjugate and parity-conserving dimension-eight effective operators by assuming the $\mathrm{SU}(2)\times\mathrm{U}(1)$ symmetry of the EW gauge field, including a Higgs doublet to incorporate the presence of an SM Higgs boson.
The presence of nonzero aQGCs would enhance the production of events with large $\wzg$ mass.
The operators that can be constrained using the $\wzg$ channel are those containing only the $\mathrm{SU}(2)$ field strength and the $\mathrm{U}(1)$ field strength, $\mathcal{L}_{\mathrm{T,0}} - \mathcal{L}_{\mathrm{T,7}}$. The Wilson coefficient of the operator $\mathcal{L}_{\mathrm{T,X}}$ is denoted by $F_{\mathrm{T,X}}/\Lambda^4$, where $\Lambda$ is the unknown scale of BSM physics. 

A simulation that includes the effects of the nonzero aQGCs in addition to the SM $\wzg$ process is performed, including the interference between the two.
To utilize the sensitivity of the invariant mass of the triboson system, the invariant mass of the three leptons and the photon, $\mllla$, is used for the aQGCs study.
In order to obtain a continuous prediction for the aQGCs signal as a function of the anomalous coupling strength, a quadratic function is used to fit the ratio of aQGC and SM yields in the last bin of the $\mllla$ distribution.
The one-dimensional 95\% confidence level (\CL) limits are calculated using the asymptotic properties of the profile likelihood ratio, using the procedure described in Section~3.2 of Ref.~\cite{Khachatryan:2014jba}, with all the other aQGC parameters, except for the coefficient being probed, set to zero.
No significant excess of events with respect to the SM prediction is observed, and the 95\% \CL observed and expected limits are summarized in Table~\ref{tab:aQGC}.
A unitarity bound is defined as the scattering energy at which the aQGC coupling strength set equal to the observed limit would result in a scattering amplitude that violates unitarity.
The value of the unitarity bound is determined using the \textsc{vbfnlo}~2.7.1 framework~\cite{VBFNLO}.
These constraints are of comparable order to those from other analyses~\cite{CMS-SMP-19-013,ZGjj_fullrun2}.

\begin{table*}[htb]
\centering
  \topcaption{Exclusion limits at the 95\% \CL for each aQGC coefficient, assuming all other coefficients are set to zero. Unitarity bounds corresponding to each operator are also listed.}
  \begin{scotch}{cccc}
  Operators & Observed limits [$\TeVns^{-4}$] & Expected limits [$\TeVns^{-4}$] & Unitarity bound [$\TeVns$]\\
  \hline
$F_{\mathrm{T,0}}/\Lambda^{4}$ & [-2.60, 2.60] & [-2.52, 2.52] & 1.32 \\
$F_{\mathrm{T,1}}/\Lambda^{4}$ & [-3.28, 3.24] & [-3.18, 3.14] & 1.48 \\
$F_{\mathrm{T,2}}/\Lambda^{4}$ & [-7.15, 7.05] & [-6.95, 6.85] & 1.35 \\
$F_{\mathrm{T,5}}/\Lambda^{4}$ & [-2.54, 2.56] & [-2.46, 2.50] & 1.55 \\
$F_{\mathrm{T,6}}/\Lambda^{4}$ & [-3.18, 3.22] & [-3.08, 3.14] & 1.61 \\
$F_{\mathrm{T,7}}/\Lambda^{4}$ & [-6.85, 7.05] & [-6.65, 6.85] & 1.71 \\
  \end{scotch}\label{tab:aQGC}
\end{table*}

\subsection{Limits on photophobic axion-like particle models}\label{sec:alp}
\begin{figure}[htb]
\centering
      \includegraphics[width=0.49\textwidth]{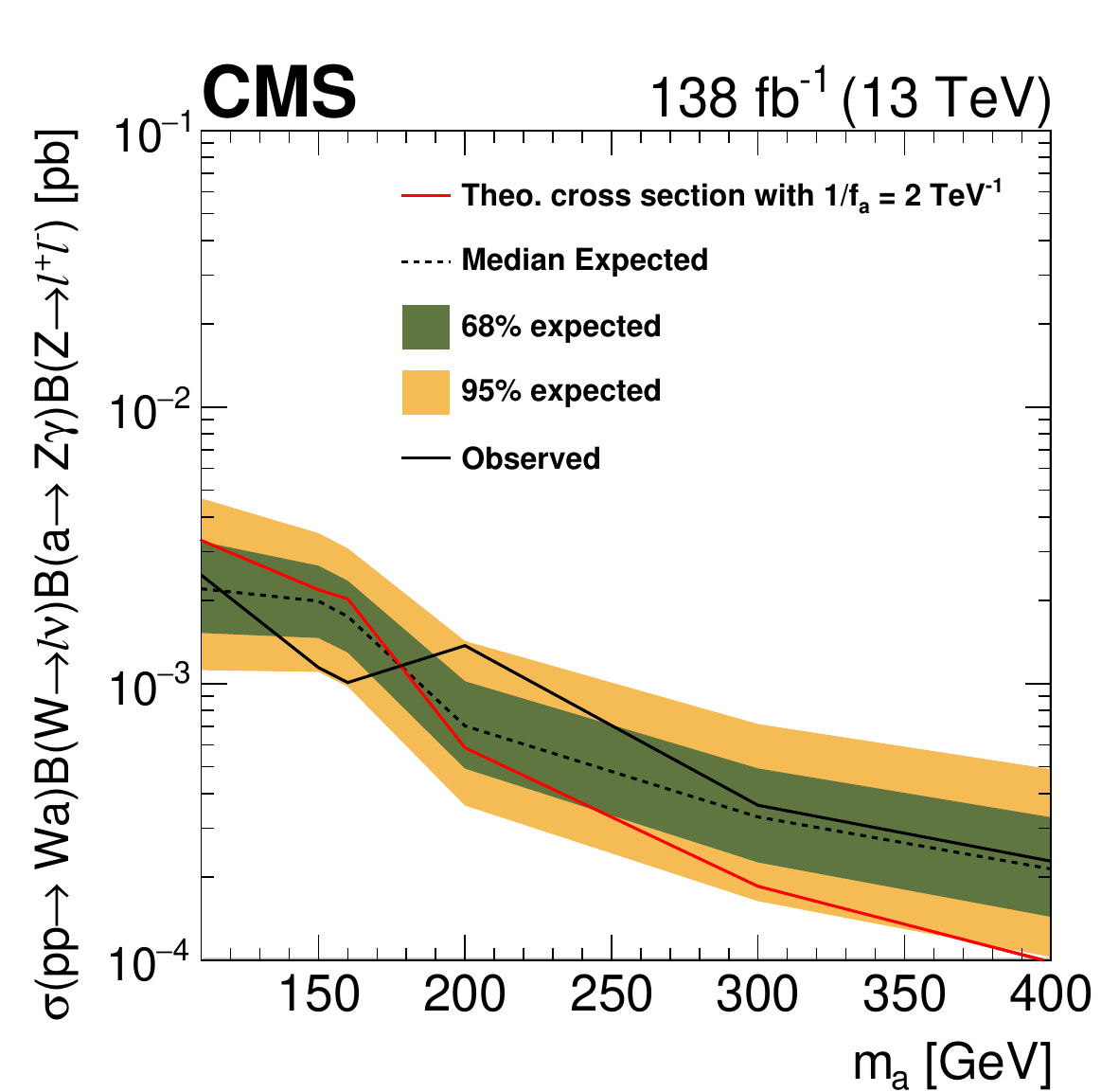}
      \includegraphics[width=0.49\textwidth]{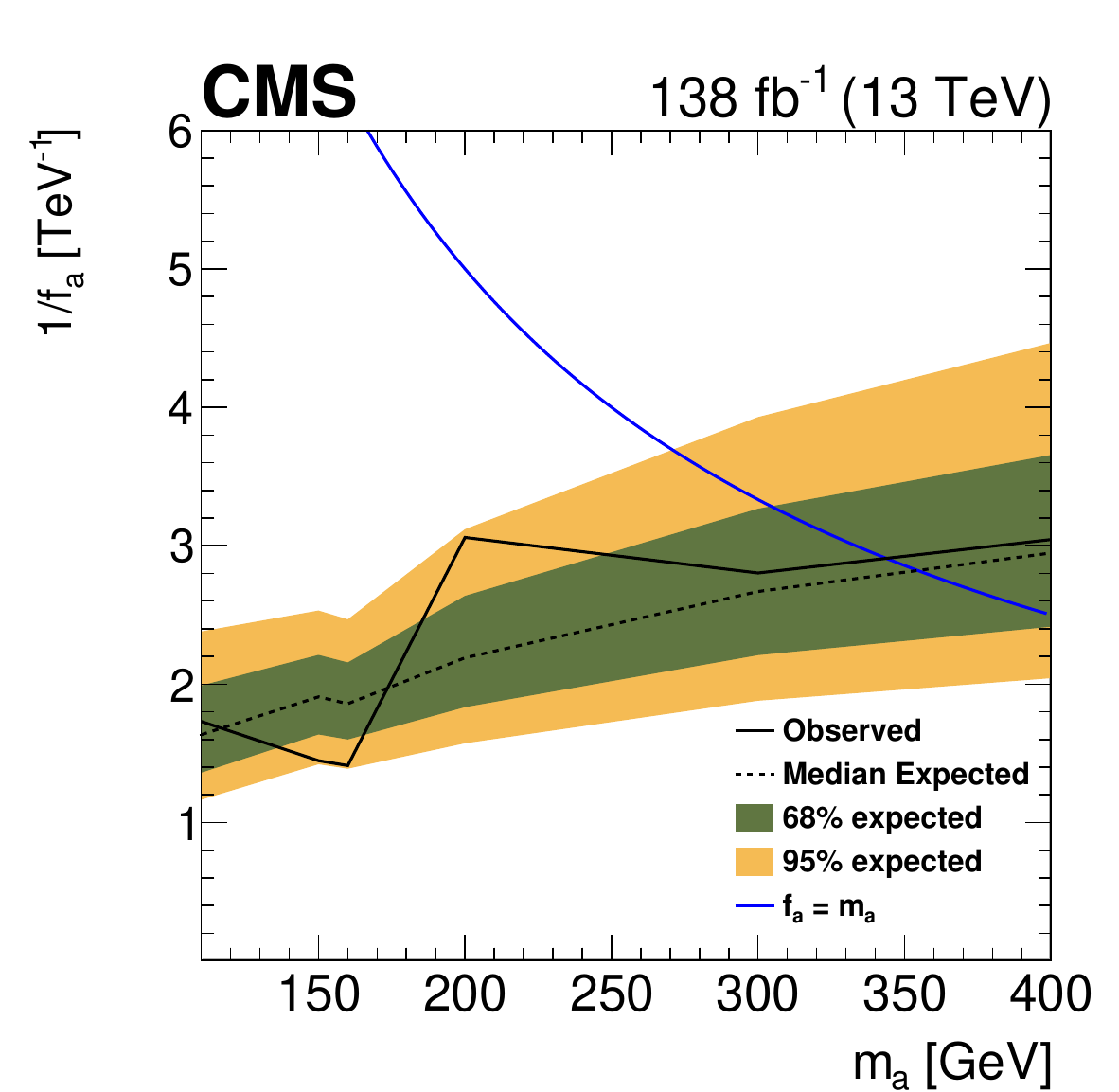}
\caption{Expected and observed 95\% upper limits on the product of the cross section and branching fraction $\sigma(\Pp\Pp\to \PW{\ALP})\mathcal{B}(\PW\to \Pell^{+}\PGn_{\Pell})\mathcal{B}(\ALP\to \PZ\PGg)\mathcal{B}(\PZ\to \Pell^{+} \Pell^{-})$ as a function of the ALP mass from 110 to 400\GeV (\cmsLeft). The red line corresponds to the theoretical prediction for $1/f_\ALP=2 \TeV^{-1}$. Expected and observed 95\% upper limits on the photophobic ALP model parameter $1/f_\ALP$ as a function of ALP mass reinterpreted from $1/f_\ALP=2\TeV^{-1}$ (\cmsRight). The blue line indicates the point at which the energy scale of $f_\ALP$ matches that of the ALP mass. The model may not be valid in the region where $m_\ALP > f_\ALP$, as discussed in Ref.~\cite{ALP_future}.}\label{fig:ALP_limit}
\end{figure}

In this analysis, a specific model called photophobic ALPs was studied~\cite{Craig:2018kne,CMS-SMP-17-013}, whose leading couplings to SM EW gauge bosons do not involve a pair of photons.
In this scenario, the $\wzg$ final state results from the production of $\PW\ALP$ followed by $\ALP\to\PZ\PGg$.
The invariant mass, $\mlla$, is used for the ALPs search.
In this case, the SM $\wzg$ contribution is regarded as a background whose normalization is determined in the simultaneous fit.
Since no significant excess of events relative to the SM prediction is observed, the \CLs criterion~\cite{Junk:1999kv,CLS2} 
 with the asymptotic approximation~\cite{Cowan:2010js} is used to evaluate 95\% CL upper limits.
The results are shown in Fig.~\ref{fig:ALP_limit}.

There is only one free parameter $f_\ALP$ that directly determines the couplings between ALPs and vector bosons, which subsequently affects the cross section times branching fraction in the case with $\sigma\mathcal{B} \propto 1/f_\ALP^2$. The limits in Fig.~\ref{fig:ALP_limit} (\cmsLeft) can be directly reinterpreted as limits in the $1/f_\ALP - m_{\ALP}$ plane. The limits on $1/f_\ALP$ are shown in Fig.~\ref{fig:ALP_limit}~(\cmsRight), providing some of the tightest limits for mass values in the range of $m_{\ALP}=200$--$400\GeV$, and also offering the first constraints set on this model for masses between $m_{\ALP}=110$--$200\GeV$ compared with the previous results in Ref.~\cite{CMS-SMP-17-013}.

\section{Summary}\label{sec:summary}
A measurement of the production of $\wzg$ with both \PW and \PZ bosons decaying leptonically has been presented.
Results are based on the data collected in proton-proton ($\Pp\Pp$) collisions at $\sqrt{s}=13\TeV$ by the CMS detector during 2016--2018, corresponding to an integrated luminosity of 138\fbinv.
Events are selected by requiring an identified photon, missing transverse momentum, as well as three identified leptons, of which two correspond to an on-shell \PZ boson.
The observed significance for the standard model (SM) signal is 5.4 standard deviations, while a significance of 3.8 standard deviations is expected based on the SM prediction.
The measured fiducial cross section of leptonic $\wzg$ production is $\sigma_{\Pp\Pp\to\Pellpm\PGn\Pellp\Pellm\PGg}=5.48 \pm 1.11\unit{fb}$, with prompt $\Pell = \Pe$ or $\PGm$, which is compatible with the prediction of $3.69 \pm 0.24\unit{fb}$ at next-to-leading order in quantum chromodynamics.
Constraints are placed on anomalous quartic gauge couplings in terms of dimension-eight operators in effective field theory, which are of comparable order to analyses of other final states.
Upper limits on the production cross section times branching fraction of photophobic axion-like particles (ALPs), denoted \ALP, are set as a function of the ALP mass. 
Equivalent limits for the ALP mass and coupling parameters are reported, including some of the most stringent constraints for mass values between $m_{\ALP}=200$--$400\GeV$, as well as the first interpretation for masses between $m_{\ALP}=110$--$200\GeV$.

\begin{acknowledgments}
  We congratulate our colleagues in the CERN accelerator departments for the excellent performance of the LHC and thank the technical and administrative staffs at CERN and at other CMS institutes for their contributions to the success of the CMS effort. In addition, we gratefully acknowledge the computing centers and personnel of the Worldwide LHC Computing Grid and other centers for delivering so effectively the computing infrastructure essential to our analyses. Finally, we acknowledge the enduring support for the construction and operation of the LHC, the CMS detector, and the supporting computing infrastructure provided by the following funding agencies: SC (Armenia), BMBWF and FWF (Austria); FNRS and FWO (Belgium); CNPq, CAPES, FAPERJ, FAPERGS, and FAPESP (Brazil); MES and BNSF (Bulgaria); CERN; CAS, MoST, and NSFC (China); MINCIENCIAS (Colombia); MSES and CSF (Croatia); RIF (Cyprus); SENESCYT (Ecuador); ERC PRG, RVTT3 and MoER TK202 (Estonia); Academy of Finland, MEC, and HIP (Finland); CEA and CNRS/IN2P3 (France); SRNSF (Georgia); BMBF, DFG, and HGF (Germany); GSRI (Greece); NKFIH (Hungary); DAE and DST (India); IPM (Iran); SFI (Ireland); INFN (Italy); MSIP and NRF (Republic of Korea); MES (Latvia); LMTLT (Lithuania); MOE and UM (Malaysia); BUAP, CINVESTAV, CONACYT, LNS, SEP, and UASLP-FAI (Mexico); MOS (Montenegro); MBIE (New Zealand); PAEC (Pakistan); MES and NSC (Poland); FCT (Portugal);  MESTD (Serbia); MICIU/AEI and PCTI (Spain); MOSTR (Sri Lanka); Swiss Funding Agencies (Switzerland); MST (Taipei); MHESI and NSTDA (Thailand); TUBITAK and TENMAK (Turkey); NASU (Ukraine); STFC (United Kingdom); DOE and NSF (USA).
   
  \hyphenation{Rachada-pisek} Individuals have received support from the Marie-Curie program and the European Research Council and Horizon 2020 Grant, contract Nos.\ 675440, 724704, 752730, 758316, 765710, 824093, 101115353, 101002207, and COST Action CA16108 (European Union); the Leventis Foundation; the Alfred P.\ Sloan Foundation; the Alexander von Humboldt Foundation; the Science Committee, project no. 22rl-037 (Armenia); the Fonds pour la Formation \`a la Recherche dans l'Industrie et dans l'Agriculture (FRIA-Belgium); the Beijing Municipal Science \& Technology Commission, No. Z191100007219010 and Fundamental Research Funds for the Central Universities (China); the Ministry of Education, Youth and Sports (MEYS) of the Czech Republic; the Shota Rustaveli National Science Foundation, grant FR-22-985 (Georgia); the Deutsche Forschungsgemeinschaft (DFG), among others, under Germany's Excellence Strategy -- EXC 2121 ``Quantum Universe" -- 390833306, and under project number 400140256 - GRK2497; the Hellenic Foundation for Research and Innovation (HFRI), Project Number 2288 (Greece); the Hungarian Academy of Sciences, the New National Excellence Program - \'UNKP, the NKFIH research grants K 131991, K 133046, K 138136, K 143460, K 143477, K 146913, K 146914, K 147048, 2020-2.2.1-ED-2021-00181, TKP2021-NKTA-64, and 2021-4.1.2-NEMZ\_KI-2024-00036 (Hungary); the Council of Science and Industrial Research, India; ICSC -- National Research Center for High Performance Computing, Big Data and Quantum Computing and FAIR -- Future Artificial Intelligence Research, funded by the NextGenerationEU program (Italy); the Latvian Council of Science; the Ministry of Education and Science, project no. 2022/WK/14, and the National Science Center, contracts Opus 2021/41/B/ST2/01369 and 2021/43/B/ST2/01552 (Poland); the Funda\c{c}\~ao para a Ci\^encia e a Tecnologia, grant CEECIND/01334/2018 (Portugal); the National Priorities Research Program by Qatar National Research Fund; MICIU/AEI/10.13039/501100011033, ERDF/EU, "European Union NextGenerationEU/PRTR", and Programa Severo Ochoa del Principado de Asturias (Spain); the Chulalongkorn Academic into Its 2nd Century Project Advancement Project, and the National Science, Research and Innovation Fund via the Program Management Unit for Human Resources \& Institutional Development, Research and Innovation, grant B39G670016 (Thailand); the Kavli Foundation; the Nvidia Corporation; the SuperMicro Corporation; the Welch Foundation, contract C-1845; and the Weston Havens Foundation (USA).

\end{acknowledgments}

\bibliography{auto_generated}

\providecommand{\href}[2]{#2}\begingroup\raggedright\begin{thebibliography}{10}%
\makeatletter
\providecommand{\hrefCMSnoop }[0]{\@secondoftwo}%
\makeatother
\providecommand{\doi}{\texttt{doi:}\begingroup \urlstyle{tt}\Url}

\bibitem{Aad:2012tfa}
\hrefCMSnoop {}{{ATLAS Collaboration}, ``Observation of a new particle in the
  search for the standard model higgs boson with the {ATLAS} detector at the
  {LHC}'',} \textit{ Phys. Lett. B} \textbf{ 716} (2012) 1,
  \href{http://dx.doi.org/10.1016/j.physletb.2012.08.020}{\doi{10.1016/j.physletb.2012.08.020}},
  \href{http://www.arXiv.org/abs/1207.7214}{\texttt{arXiv:1207.7214}}.

\bibitem{Chatrchyan:2012ufa}
\hrefCMSnoop {}{{CMS Collaboration}, ``Observation of a new boson at a mass of
  125 {GeV} with the {CMS} experiment at the {LHC}'',} \textit{ Phys. Lett. B}
  \textbf{ 716} (2012) 30,
  \href{http://dx.doi.org/10.1016/j.physletb.2012.08.021}{\doi{10.1016/j.physletb.2012.08.021}},
\href{http://www.arXiv.org/abs/1207.7235}{\texttt{arXiv:1207.7235}}.

\bibitem{Chatrchyan:2012ufa_long}
\hrefCMSnoop {}{{CMS Collaboration}, ``Observation of a new boson with mass
  near 125 {GeV} in {$\Pp\Pp$} collisions at $\sqrt{s} =$ 7 and 8 {TeV}'',}
  \textit{ JHEP} \textbf{ 06} (2013) 081,
  \href{http://dx.doi.org/10.1007/JHEP06(2013)081}{\doi{10.1007/JHEP06(2013)081}},
\href{http://www.arXiv.org/abs/1303.4571}{\texttt{arXiv:1303.4571}}.

\bibitem{PhysRevLett.13.321}
\hrefCMSnoop {}{F.~Englert and R.~Brout, ``Broken symmetry and the mass of
  gauge vector mesons'',} \textit{ Phys. Rev. Lett.} \textbf{ 13} (1964) 321,
  \href{http://dx.doi.org/10.1103/PhysRevLett.13.321}{\doi{10.1103/PhysRevLett.13.321}}.

\bibitem{PhysRevLett.13.508}
\hrefCMSnoop {}{P.~W. Higgs, ``Broken symmetries and the masses of gauge
  bosons'',} \textit{ Phys. Rev. Lett.} \textbf{ 13} (1964) 508,
  \href{http://dx.doi.org/10.1103/PhysRevLett.13.508}{\doi{10.1103/PhysRevLett.13.508}}.

\bibitem{PhysRevLett.13.585}
\hrefCMSnoop {}{G.~S. Guralnik, C.~R. Hagen, and T.~W.~B. Kibble, ``Global
  conservation laws and massless particles'',} \textit{ Phys. Rev. Lett.}
  \textbf{ 13} (1964) 585,
  \href{http://dx.doi.org/10.1103/PhysRevLett.13.585}{\doi{10.1103/PhysRevLett.13.585}}.

\bibitem{CMS:2024phk}
\hrefCMSnoop {}{{CMS Collaboration}, ``Searches for {Higgs} boson production
  through decays of heavy resonances'',} \textit{ Phys. Rept.} \textbf{ 1115}
  (2025) 368,
  \href{http://dx.doi.org/10.1016/j.physrep.2024.09.004}{\doi{10.1016/j.physrep.2024.09.004}},
  \href{http://www.arXiv.org/abs/2403.16926}{\texttt{arXiv:2403.16926}}.

\bibitem{ATLAS:2024lda}
\hrefCMSnoop {}{{ATLAS Collaboration}, ``The quest to discover supersymmetry at
  the {ATLAS} experiment'',} 2024.
  \href{http://www.arXiv.org/abs/2403.02455}{\texttt{arXiv:2403.02455}}.
  Submitted to \textit{Phys. Rep.}

\bibitem{ATLAS:2024fdw}
\hrefCMSnoop {}{{ATLAS Collaboration}, ``Exploration at the high-energy
  frontier: {ATLAS} {Run} 2 searches investigating the exotic jungle beyond the
  {Standard Model}'',} 2024.
  \href{http://www.arXiv.org/abs/2403.09292}{\texttt{arXiv:2403.09292}}.
  Submitted to \textit{Phys. Rep.}

\bibitem{CMS-SMP-23-004}
\hrefCMSnoop {}{{CMS Collaboration}, ``Stairway to discovery: a report on the
  {CMS} programme of cross section measurements from millibarns to
  femtobarns'',} \textit{ Phys. Rept.} \textbf{ 1115} (2025) 3,
  \href{http://dx.doi.org/10.1016/j.physrep.2024.11.005}{\doi{10.1016/j.physrep.2024.11.005}},
  \href{http://www.arXiv.org/abs/2405.18661}{\texttt{arXiv:2405.18661}}.

\bibitem{CERN-EP-2024-093}
\href {https://arxiv.org/abs/2404.06829}{{ATLAS Collaboration}, ``Electroweak,
  {QCD} and flavour physics studies with {ATLAS} data from {Run} 2 of the
  {LHC}'',} 2024.
  \href{http://www.arXiv.org/abs/2404.06829}{\texttt{arXiv:2404.06829}}.
  Submitted to \textit{Phys. Rep.}

\bibitem{CMS-SMP-19-014}
\hrefCMSnoop {}{{CMS Collaboration}, ``Observation of the production of three
  massive gauge bosons at {$\sqrt{s}=13\,\mathrm{TeV}$}'',} \textit{ Phys. Rev.
  Lett.} \textbf{ 125} (2020) 151802,
  \href{http://dx.doi.org/10.1103/PhysRevLett.125.151802}{\doi{10.1103/PhysRevLett.125.151802}},
  \href{http://www.arXiv.org/abs/2006.11191}{\texttt{arXiv:2006.11191}}.

\bibitem{ATLAS-WGG}
\hrefCMSnoop {}{{ATLAS Collaboration}, ``Observation of {$\PW\PGg\PGg$}
  triboson production in proton-proton collisions at
  {$\sqrt{s}=13\,\mathrm{TeV}$} with the {ATLAS} detector'',} \textit{ Phys.
  Lett. B} \textbf{ 848} (2024) 138400,
  \href{http://dx.doi.org/10.1016/j.physletb.2023.138400}{\doi{10.1016/j.physletb.2023.138400}},
  \href{http://www.arXiv.org/abs/2308.03041}{\texttt{arXiv:2308.03041}}.

\bibitem{CMS-SMP-19-013}
\hrefCMSnoop {}{{CMS Collaboration}, ``Measurements of the
  {$\Pp\Pp\rightarrow\PW^{\pm}\PGg\PGg$} and {$\Pp\Pp\rightarrow\PZ\PGg\PGg$}
  cross sections at {$\sqrt{s}=13\,\mathrm{TeV}$} and limits on anomalous
  quartic gauge couplings'',} \textit{ JHEP} \textbf{ 10} (2021) 174,
  \href{http://dx.doi.org/10.1007/JHEP10(2021)174}{\doi{10.1007/JHEP10(2021)174}},
  \href{http://www.arXiv.org/abs/2105.12780}{\texttt{arXiv:2105.12780}}.

\bibitem{CMS-SMP-22-006}
\hrefCMSnoop {}{{CMS Collaboration}, ``Observation of {$\PW\PW\PGg$} production
  and search for {H}$\ensuremath{\gamma}$ production in proton-proton
  collisions at {$\sqrt{s}=13\text{ }\text{ }\mathrm{TeV}$}'',} \textit{ Phys.
  Rev. Lett.} \textbf{ 132} (2024) 121901,
  \href{http://dx.doi.org/10.1103/PhysRevLett.132.121901}{\doi{10.1103/PhysRevLett.132.121901}},
  \href{http://www.arXiv.org/abs/2310.05164}{\texttt{arXiv:2310.05164}}.

\bibitem{PhysRevLett.132.021802}
\hrefCMSnoop {}{{ATLAS Collaboration}, ``Observation of
  {$WZ\ensuremath{\gamma}$} production in {$\Pp\Pp$} collisions at
  $\sqrt{s}=13\text{ }\text{ }\mathrm{TeV}$ with the {ATLAS} detector'',}
  \textit{ Phys. Rev. Lett.} \textbf{ 132} (2024) 021802,
  \href{http://dx.doi.org/10.1103/PhysRevLett.132.021802}{\doi{10.1103/PhysRevLett.132.021802}},
\href{http://www.arXiv.org/abs/2305.16994}{\texttt{arXiv:2305.16994}}.

\bibitem{paper_aqgc}
C.~Degrande\hrefCMSnoop {}{ { et~al.}, ``Effective field theory: A modern
  approach to anomalous couplings'',} \textit{ Annals Phys.} \textbf{ 335}
  (2013) 21,
  \href{http://dx.doi.org/10.1016/j.aop.2013.04.016}{\doi{10.1016/j.aop.2013.04.016}},
  \href{http://www.arXiv.org/abs/1205.4231}{\texttt{arXiv:1205.4231}}.

\bibitem{Weinberg:ALP}
\hrefCMSnoop {}{S.~Weinberg, ``A new light boson?'',} \textit{ Phys. Rev.
  Lett.} \textbf{ 40} (1978) 223,
  \href{http://dx.doi.org/10.1103/PhysRevLett.40.223}{\doi{10.1103/PhysRevLett.40.223}}.

\bibitem{Wilczek:ALP}
\hrefCMSnoop {}{F.~Wilczek, ``Problem of strong $p$ and $t$ invariance in the
  presence of instantons'',} \textit{ Phys. Rev. Lett.} \textbf{ 40} (1978)
  279,
  \href{http://dx.doi.org/10.1103/PhysRevLett.40.279}{\doi{10.1103/PhysRevLett.40.279}}.

\bibitem{ALP1}
\hrefCMSnoop {}{K.~Choi, S.~H. Im, and C.~Sub~Shin, ``Recent progress in the
  physics of axions and axion-like particles'',} \textit{ Ann. Rev. Nucl. Part.
  Sci.} \textbf{ 71} (2021) 225,
  \href{http://dx.doi.org/10.1146/annurev-nucl-120720-031147}{\doi{10.1146/annurev-nucl-120720-031147}},
  \href{http://www.arXiv.org/abs/2012.05029}{\texttt{arXiv:2012.05029}}.

\bibitem{ALP_atlas}
\hrefCMSnoop {}{{ATLAS Collaboration}, ``Measurement of light-by-light
  scattering and search for axion-like particles with 2.2 nb$^{-1}$ of {Pb+Pb}
  data with the {ATLAS} detector'',} \textit{ JHEP} \textbf{ 03} (2021) 243,
  \href{http://dx.doi.org/10.1007/JHEP03(2021)243}{\doi{10.1007/JHEP03(2021)243}},
  \href{http://www.arXiv.org/abs/2008.05355}{\texttt{arXiv:2008.05355}}.
  [Erratum: \DOI{10.1007/JHEP11(2021)050}].

\bibitem{ALP_future}
\hrefCMSnoop {}{M.~Bauer, M.~Heiles, M.~Neubert, and A.~Thamm, ``Axion-like
  particles at future colliders'',} \textit{ Eur. Phys. J. C} \textbf{ 79}
  (2019) 74,
  \href{http://dx.doi.org/10.1140/epjc/s10052-019-6587-9}{\doi{10.1140/epjc/s10052-019-6587-9}},
  \href{http://www.arXiv.org/abs/1808.10323}{\texttt{arXiv:1808.10323}}.

\bibitem{DiLuzio:ALP}
\hrefCMSnoop {}{L.~Di~Luzio, M.~Giannotti, E.~Nardi, and L.~Visinelli, ``The
  landscape of {QCD} axion models'',} \textit{ Phys. Rept.} \textbf{ 870}
  (2020) 1,
  \href{http://dx.doi.org/10.1016/j.physrep.2020.06.002}{\doi{10.1016/j.physrep.2020.06.002}},
  \href{http://www.arXiv.org/abs/2003.01100}{\texttt{arXiv:2003.01100}}.

\bibitem{Axion1}
\hrefCMSnoop {}{R.~D. Peccei and H.~R. Quinn, ``$\mathrm{CP}$ conservation in
  the presence of pseudoparticles'',} \textit{ Phys. Rev. Lett.} \textbf{ 38}
  (1977) 1440,
  \href{http://dx.doi.org/10.1103/PhysRevLett.38.1440}{\doi{10.1103/PhysRevLett.38.1440}}.

\bibitem{Axion2}
\hrefCMSnoop {}{R.~D. Peccei and H.~R. Quinn, ``Constraints imposed by
  $\mathrm{CP}$ conservation in the presence of pseudoparticles'',} \textit{
  Phys. Rev. D} \textbf{ 16} (1977) 1791,
  \href{http://dx.doi.org/10.1103/PhysRevD.16.1791}{\doi{10.1103/PhysRevD.16.1791}}.

\bibitem{Craig:2018kne}
\hrefCMSnoop {}{N.~Craig, A.~Hook, and S.~Kasko, ``The photophobic {ALP}'',}
  \textit{ JHEP} \textbf{ 09} (2018) 028,
  \href{http://dx.doi.org/10.1007/JHEP09(2018)028}{\doi{10.1007/JHEP09(2018)028}},
  \href{http://www.arXiv.org/abs/1805.06538}{\texttt{arXiv:1805.06538}}.

\bibitem{hepdata}
\hrefCMSnoop {}{}{HEPData} record for this analysis, 2025.
\newblock
  \href{http://dx.doi.org/10.17182/hepdata.157601}{\doi{10.17182/hepdata.157601}}.

\bibitem{Chatrchyan:2008zzk}
\hrefCMSnoop {}{{CMS Collaboration}, ``The {CMS} experiment at the {CERN}
  {LHC}'',} \textit{ JINST} \textbf{ 3} (2008) S08004,
  \href{http://dx.doi.org/10.1088/1748-0221/3/08/S08004}{\doi{10.1088/1748-0221/3/08/S08004}}.

\bibitem{trigger_system}
\hrefCMSnoop {}{{CMS Collaboration}, ``The {CMS} trigger system'',} \textit{
  JINST} \textbf{ 12} (2017) P01020,
  \href{http://dx.doi.org/10.1088/1748-0221/12/01/P01020}{\doi{10.1088/1748-0221/12/01/P01020}},
  \href{http://www.arXiv.org/abs/1609.02366}{\texttt{arXiv:1609.02366}}.

\bibitem{MGatNLO}
J.~Alwall\hrefCMSnoop {}{ { et~al.}, ``The automated computation of tree-level
  and next-to-leading order differential cross sections, and their matching to
  parton shower simulations'',} \textit{ JHEP} \textbf{ 07} (2014) 079,
  \href{http://dx.doi.org/10.1007/JHEP07(2014)079}{\doi{10.1007/JHEP07(2014)079}},
\href{http://www.arXiv.org/abs/1405.0301}{\texttt{arXiv:1405.0301}}.

\bibitem{Frederix:2012ps}
\hrefCMSnoop {}{R.~Frederix and S.~Frixione, ``Merging meets matching in
  {MC@NLO}'',} \textit{ JHEP} \textbf{ 12} (2012) 061,
  \href{http://dx.doi.org/10.1007/JHEP12(2012)061}{\doi{10.1007/JHEP12(2012)061}},
\href{http://www.arXiv.org/abs/1209.6215}{\texttt{arXiv:1209.6215}}.

\bibitem{powheg_tw}
\hrefCMSnoop {}{E.~Re, ``Single-top {W}t-channel production matched with parton
  showers using the {POWHEG} method'',} \textit{ Eur. Phys. J. C} \textbf{ 71}
  (2011) 1547,
  \href{http://dx.doi.org/10.1140/epjc/s10052-011-1547-z}{\doi{10.1140/epjc/s10052-011-1547-z}},
  \href{http://www.arXiv.org/abs/1009.2450}{\texttt{arXiv:1009.2450}}.

\bibitem{POWHEG1}
\hrefCMSnoop {}{P.~Nason, ``A new method for combining {NLO QCD} with shower
  {Monte Carlo} algorithms'',} \textit{ JHEP} \textbf{ 11} (2004) 040,
  \href{http://dx.doi.org/10.1088/1126-6708/2004/11/040}{\doi{10.1088/1126-6708/2004/11/040}},
\href{http://www.arXiv.org/abs/hep-ph/0409146}{\texttt{arXiv:hep-ph/0409146}}.

\bibitem{POWHEG2}
\hrefCMSnoop {}{S.~Frixione, P.~Nason, and C.~Oleari, ``Matching {NLO QCD}
  computations with parton shower simulations: the {POWHEG} method'',} \textit{
  JHEP} \textbf{ 11} (2007) 070,
  \href{http://dx.doi.org/10.1088/1126-6708/2007/11/070}{\doi{10.1088/1126-6708/2007/11/070}},
\href{http://www.arXiv.org/abs/0709.2092}{\texttt{arXiv:0709.2092}}.

\bibitem{POWHEG3}
\hrefCMSnoop {}{S.~Alioli, P.~Nason, C.~Oleari, and E.~Re, ``A general
  framework for implementing {NLO} calculations in shower {Monte Carlo}
  programs: the {POWHEG} box'',} \textit{ JHEP} \textbf{ 06} (2010) 043,
  \href{http://dx.doi.org/10.1007/JHEP06(2010)043}{\doi{10.1007/JHEP06(2010)043}},
\href{http://www.arXiv.org/abs/1002.2581}{\texttt{arXiv:1002.2581}}.

\bibitem{powheg_VV}
\hrefCMSnoop {}{T.~Melia, P.~Nason, R.~Rontsch, and G.~Zanderighi,
  ``$\text{W}^{+}\text{W}^{-}$, {\PW}{\PZ} and {\PZ}{\PZ} production in the
  {POWHEG BOX}'',} \textit{ JHEP} \textbf{ 11} (2011) 078,
  \href{http://dx.doi.org/10.1007/JHEP11(2011)078}{\doi{10.1007/JHEP11(2011)078}},
\href{http://www.arXiv.org/abs/1107.5051}{\texttt{arXiv:1107.5051}}.

\bibitem{MCFM}
\hrefCMSnoop {}{J.~M. Campbell and R.~K. Ellis, ``{MCFM} for the tevatron and
  the {LHC}'',} \textit{ Nucl. Phys. Proc. Suppl.} \textbf{ 205-206} (2010) 10,
  \href{http://dx.doi.org/10.1016/j.nuclphysbps.2010.08.011}{\doi{10.1016/j.nuclphysbps.2010.08.011}},
\href{http://www.arXiv.org/abs/1007.3492}{\texttt{arXiv:1007.3492}}.

\bibitem{mg5:reweight}
\hrefCMSnoop {}{O.~Mattelaer, ``On the maximal use of {Monte Carlo} samples:
  re-weighting events at {NLO} accuracy'',} \textit{ Eur. Phys. J. C} \textbf{
  76} (2016) 674,
  \href{http://dx.doi.org/10.1140/epjc/s10052-016-4533-7}{\doi{10.1140/epjc/s10052-016-4533-7}},
\href{http://www.arXiv.org/abs/1607.00763}{\texttt{arXiv:1607.00763}}.

\bibitem{CP5}
\hrefCMSnoop {}{{CMS Collaboration}, ``Extraction and validation of a new set
  of {CMS} {PYTHIA8} tunes from underlying-event measurements'',} \textit{ Eur.
  Phys. J. C} \textbf{ 80} (2020) 4,
  \href{http://dx.doi.org/10.1140/epjc/s10052-019-7499-4}{\doi{10.1140/epjc/s10052-019-7499-4}},
  \href{http://www.arXiv.org/abs/1903.12179}{\texttt{arXiv:1903.12179}}.

\bibitem{NNPDF:2017mvq}
\hrefCMSnoop {}{{NNPDF} Collaboration, ``Parton distributions from
  high-precision collider data'',} \textit{ Eur. Phys. J. C} \textbf{ 77}
  (2017) 663,
  \href{http://dx.doi.org/10.1140/epjc/s10052-017-5199-5}{\doi{10.1140/epjc/s10052-017-5199-5}},
  \href{http://www.arXiv.org/abs/1706.00428}{\texttt{arXiv:1706.00428}}.

\bibitem{geant4}
\hrefCMSnoop {}{{GEANT4} Collaboration, ``{\GEANTfour}---a simulation
  toolkit'',} \textit{ Nucl. Instrum. Meth. A} \textbf{ 506} (2003) 250,
\href{http://dx.doi.org/10.1016/S0168-9002(03)01368-8}{\doi{10.1016/S0168-9002(03)01368-8}}.

\bibitem{geant4_2}
\hrefCMSnoop {}{J.~Allison { et~al.}, ``{GEANT4} developments and
  applications'',} \textit{ IEEE Trans. Nucl. Sci.} \textbf{ 53} (2006) 270,
  \href{http://dx.doi.org/10.1109/TNS.2006.869826}{\doi{10.1109/TNS.2006.869826}}.

\bibitem{paticleflow}
\hrefCMSnoop {}{{CMS Collaboration}, ``Particle-flow reconstruction and global
  event description with the {CMS} detector'',} \textit{ JINST} \textbf{ 12}
  (2017) P10003,
  \href{http://dx.doi.org/10.1088/1748-0221/12/10/P10003}{\doi{10.1088/1748-0221/12/10/P10003}},
\href{http://www.arXiv.org/abs/1706.04965}{\texttt{arXiv:1706.04965}}.

\bibitem{CMS-TDR-15-02}
\href {http://cds.cern.ch/record/2020886}{{CMS Collaboration}, ``Technical
  proposal for the {Phase-II} upgrade of the {Compact Muon Solenoid}'',} CMS
  Technical Proposal CERN-LHCC-2015-010, CMS-TDR-15-02, 2015.

\bibitem{Sirunyan_2018}
\hrefCMSnoop {}{{CMS Collaboration}, ``Performance of the {CMS} muon detector
  and muon reconstruction with proton-proton collisions at $\sqrt{s}=$ 13
  {TeV}'',} \textit{ JINST} \textbf{ 13} (2018) P06015,
  \href{http://dx.doi.org/10.1088/1748-0221/13/06/P06015}{\doi{10.1088/1748-0221/13/06/P06015}},
  \href{http://www.arXiv.org/abs/1804.04528}{\texttt{arXiv:1804.04528}}.

\bibitem{CMS:2019kuk}
\hrefCMSnoop {}{{CMS Collaboration}, ``Pileup mitigation at {CMS} in 13{\TeV}
  data'',} \textit{ JINST} \textbf{ 15} (2020) P09018,
  \href{http://dx.doi.org/10.1088/1748-0221/15/09/P09018}{\doi{10.1088/1748-0221/15/09/P09018}},
  \href{http://www.arXiv.org/abs/2003.00503}{\texttt{arXiv:2003.00503}}.

\bibitem{Sirunyan_2021:eg}
\hrefCMSnoop {}{{CMS Collaboration}, ``Electron and photon reconstruction and
  identification with the {CMS} experiment at the {CERN} {LHC}'',} \textit{
  JINST} \textbf{ 16} (2021) P05014,
  \href{http://dx.doi.org/10.1088/1748-0221/16/05/P05014}{\doi{10.1088/1748-0221/16/05/P05014}},
  \href{http://www.arXiv.org/abs/2012.06888}{\texttt{arXiv:2012.06888}}.

\bibitem{antikt}
\hrefCMSnoop {}{M.~Cacciari, G.~P. Salam, and G.~Soyez, ``The anti-{\kt} jet
  clustering algorithm'',} \textit{ JHEP} \textbf{ 04} (2008) 063,
  \href{http://dx.doi.org/10.1088/1126-6708/2008/04/063}{\doi{10.1088/1126-6708/2008/04/063}},
\href{http://www.arXiv.org/abs/0802.1189}{\texttt{arXiv:0802.1189}}.

\bibitem{jer}
\hrefCMSnoop {}{{CMS Collaboration}, ``Jet energy scale and resolution in the
  {CMS} experiment in {$\Pp\Pp$} collisions at 8 {TeV}'',} \textit{ JINST}
  \textbf{ 12} (2017) P02014,
  \href{http://dx.doi.org/10.1088/1748-0221/12/02/P02014}{\doi{10.1088/1748-0221/12/02/P02014}},
  \href{http://www.arXiv.org/abs/1607.03663}{\texttt{arXiv:1607.03663}}.

\bibitem{Sirunyan:2017ezt}
\hrefCMSnoop {}{{CMS Collaboration}, ``Identification of heavy-flavour jets
  with the {CMS} detector in pp collisions at 13 {TeV}'',} \textit{ JINST}
  \textbf{ 13} (2018) P05011,
  \href{http://dx.doi.org/10.1088/1748-0221/13/05/P05011}{\doi{10.1088/1748-0221/13/05/P05011}},
\href{http://www.arXiv.org/abs/1712.07158}{\texttt{arXiv:1712.07158}}.

\bibitem{CMS:2019ctu}
\hrefCMSnoop {}{{CMS Collaboration}, ``Performance of missing transverse
  momentum reconstruction in proton-proton collisions at $\sqrt{s} = 13$\,{TeV}
  using the {CMS} detector'',} \textit{ JINST} \textbf{ 14} (2019) P07004,
  \href{http://dx.doi.org/10.1088/1748-0221/14/07/P07004}{\doi{10.1088/1748-0221/14/07/P07004}},
\href{http://www.arXiv.org/abs/1903.06078}{\texttt{arXiv:1903.06078}}.

\bibitem{tagandprobe}
\hrefCMSnoop {}{{CMS Collaboration}, ``Measurement of the inclusive {\PW} and
  {\PZ} production cross sections in {$\Pp\Pp$} collisions at $\sqrt{s} =$ 7
  {TeV} with the {CMS} experiment'',} \textit{ JHEP} \textbf{ 10} (2011) 132,
  \href{http://dx.doi.org/10.1007/JHEP10(2011)132}{\doi{10.1007/JHEP10(2011)132}},
  \href{http://www.arXiv.org/abs/1107.4789}{\texttt{arXiv:1107.4789}}.

\bibitem{CMS-TRG-19-001}
\hrefCMSnoop {}{{CMS Collaboration}, ``Performance of the {CMS} high-level
  trigger during {LHC} {Run} 2'',} \textit{ JINST} \textbf{ 19} (2024) P11021,
  \href{http://dx.doi.org/10.1088/1748-0221/19/11/P11021}{\doi{10.1088/1748-0221/19/11/P11021}},
  \href{http://www.arXiv.org/abs/2410.17038}{\texttt{arXiv:2410.17038}}.

\bibitem{cms:l1_trigger}
\hrefCMSnoop {}{{CMS Collaboration}, ``Performance of the {CMS} {Level-1}
  trigger in proton-proton collisions at {$\sqrt{s} = 13\TeV$}'',} \textit{
  JINST} \textbf{ 15} (2020) P10017,
  \href{http://dx.doi.org/10.1088/1748-0221/15/10/P10017}{\doi{10.1088/1748-0221/15/10/P10017}},
  \href{http://www.arXiv.org/abs/2006.10165}{\texttt{arXiv:2006.10165}}.

\bibitem{ZGjj_fullrun2}
\hrefCMSnoop {}{{CMS Collaboration}, ``Measurement of the electroweak
  production of {Z$\gamma$} and two jets in proton-proton collisions at
  {$\sqrt{s} = 13\TeV$} and constraints on anomalous quartic gauge
  couplings'',} \textit{ Phys. Rev. D} \textbf{ 104} (2021) 072001,
  \href{http://dx.doi.org/10.1103/PhysRevD.104.072001}{\doi{10.1103/PhysRevD.104.072001}},
  \href{http://www.arXiv.org/abs/2106.11082}{\texttt{arXiv:2106.11082}}.

\bibitem{WG_fullrun2}
\hrefCMSnoop {}{{CMS Collaboration}, ``Measurement of the {$\PW\gamma$}
  production cross section in proton-proton collisions at {$\sqrt{s}=13\TeV$}
  and constraints on effective field theory coefficients'',} \textit{ Phys.
  Rev. Lett.} \textbf{ 126} (2021) 252002,
  \href{http://dx.doi.org/10.1103/PhysRevLett.126.252002}{\doi{10.1103/PhysRevLett.126.252002}},
  \href{http://www.arXiv.org/abs/2102.02283}{\texttt{arXiv:2102.02283}}.

\bibitem{CAT-23-001}
\hrefCMSnoop {}{{CMS Collaboration}, ``The {CMS} statistical analysis and
  combination tool: {\textsc{Combine}}'',} \textit{ Comput. Softw. Big Sci.}
  \textbf{ 8} (2024) 19,
  \href{http://dx.doi.org/10.1007/s41781-024-00121-4}{\doi{10.1007/s41781-024-00121-4}},
  \href{http://www.arXiv.org/abs/2404.06614}{\texttt{arXiv:2404.06614}}.

\bibitem{lumi2016}
\hrefCMSnoop {}{{CMS Collaboration}, ``Precision luminosity measurement in
  proton-proton collisions at $\sqrt{s} =$ 13 {TeV} in 2015 and 2016 at
  {CMS}'',} \textit{ Eur. Phys. J. C} \textbf{ 81} (2021) 800,
  \href{http://dx.doi.org/10.1140/epjc/s10052-021-09538-2}{\doi{10.1140/epjc/s10052-021-09538-2}},
  \href{http://www.arXiv.org/abs/2104.01927}{\texttt{arXiv:2104.01927}}.

\bibitem{lumi2017}
\href {https://cds.cern.ch/record/2621960/}{{CMS Collaboration}, ``{CMS}
  luminosity measurement for the 2017 data-taking period at $\sqrt{s}$ = 13
  {TeV}'',} CMS Physics Analysis Summary CMS-PAS-LUM-17-004, 2018.

\bibitem{lumi2018}
\href {https://cds.cern.ch/record/2676164/}{{CMS Collaboration}, ``{CMS}
  luminosity measurement for the 2018 data-taking period at $\sqrt{s}$ = 13
  {TeV}'',} CMS Physics Analysis Summary CMS-PAS-LUM-18-002, 2019.

\bibitem{Butterworth:2015oua}
\hrefCMSnoop {}{J.~Butterworth { et~al.}, ``{PDF4LHC} recommendations for {LHC
  Run II}'',} \textit{ J. Phys. G} \textbf{ 43} (2016) 023001,
  \href{http://dx.doi.org/10.1088/0954-3899/43/2/023001}{\doi{10.1088/0954-3899/43/2/023001}},
\href{http://www.arXiv.org/abs/1510.03865}{\texttt{arXiv:1510.03865}}.

\bibitem{Eboli:2006wa}
\hrefCMSnoop {}{O.~J.~P. {\'{E}}boli, M.~C. Gonzalez-Garcia, and J.~K.
  Mizukoshi, ``{$\Pp\Pp$$\rightarrow$ jje$^\pm$$\mu^\pm$$\nu\nu$} and
  {jje$^\pm\mu^\mp\nu\nu$} at {\cal o}($\alpha^6_{\rm em}$) and {\cal
  o}($\alpha_{\rm em}^4\, \alpha_{\rm s}^2$) for the study of the quartic
  electroweak gauge boson vertex at {CERN} {LHC}'',} \textit{ Phys. Rev. D}
  \textbf{ 74} (2006) 073005,
  \href{http://dx.doi.org/10.1103/PhysRevD.74.073005}{\doi{10.1103/PhysRevD.74.073005}},
\href{http://www.arXiv.org/abs/hep-ph/0606118}{\texttt{arXiv:hep-ph/0606118}}.

\bibitem{Khachatryan:2014jba}
\hrefCMSnoop {}{{CMS Collaboration}, ``Precise determination of the mass of the
  {Higgs} boson and tests of compatibility of its couplings with the standard
  model predictions using proton collisions at 7 and 8 {TeV}'',} \textit{ Eur.
  Phys. J. C} \textbf{ 75} (2015) 212,
  \href{http://dx.doi.org/10.1140/epjc/s10052-015-3351-7}{\doi{10.1140/epjc/s10052-015-3351-7}},
\href{http://www.arXiv.org/abs/1412.8662}{\texttt{arXiv:1412.8662}}.

\bibitem{VBFNLO}
\hrefCMSnoop {}{K.~Arnold { et~al.}, ``{VBFNLO}: A parton level {Monte Carlo}
  for processes with electroweak bosons'',} \textit{ Comput. Phys. Commun.}
  \textbf{ 180} (2009) 1661,
  \href{http://dx.doi.org/10.1016/j.cpc.2009.03.006}{\doi{10.1016/j.cpc.2009.03.006}},
\href{http://www.arXiv.org/abs/0811.4559}{\texttt{arXiv:0811.4559}}.

\bibitem{CMS-SMP-17-013}
\hrefCMSnoop {}{{CMS Collaboration}, ``Search for the production of
  {$\PWpm\PWpm\PWmp$} events at {$\sqrt{s} = 13\TeV$}'',} \textit{ Phys. Rev.
  D} \textbf{ 100} (2019) 012004,
  \href{http://dx.doi.org/10.1103/PhysRevD.100.012004}{\doi{10.1103/PhysRevD.100.012004}},
  \href{http://www.arXiv.org/abs/1905.04246}{\texttt{arXiv:1905.04246}}.

\bibitem{Junk:1999kv}
\hrefCMSnoop {}{T.~Junk, ``Confidence level computation for combining searches
  with small statistics'',} \textit{ Nucl. Instrum. Meth. A} \textbf{ 434}
  (1999) 435,
  \href{http://dx.doi.org/10.1016/S0168-9002(99)00498-2}{\doi{10.1016/S0168-9002(99)00498-2}},
\href{http://www.arXiv.org/abs/hep-ex/9902006}{\texttt{arXiv:hep-ex/9902006}}.

\bibitem{CLS2}
\hrefCMSnoop {}{A.~L. Read, ``Presentation of search results: The
  {CL$_{\mathrm{s}}$} technique'',} \textit{ J. Phys. G} \textbf{ 28} (2002)
  2693,
\href{http://dx.doi.org/10.1088/0954-3899/28/10/313}{\doi{10.1088/0954-3899/28/10/313}}.

\bibitem{Cowan:2010js}
\hrefCMSnoop {}{G.~Cowan, K.~Cranmer, E.~Gross, and O.~Vitells, ``Asymptotic
  formulae for likelihood-based tests of new physics'',} \textit{ Eur. Phys. J.
  C} \textbf{ 71} (2011) 1554,
  \href{http://dx.doi.org/10.1140/epjc/s10052-011-1554-0}{\doi{10.1140/epjc/s10052-011-1554-0}},
  \href{http://www.arXiv.org/abs/1007.1727}{\texttt{arXiv:1007.1727}}.
[Erratum: \DOI{10.1140/epjc/s10052-013-2501-z}].

\end{thebibliography}\endgroup
\cleardoublepage \appendix\section{The CMS Collaboration \label{app:collab}}\begin{sloppypar}\hyphenpenalty=5000\widowpenalty=500\clubpenalty=5000
\cmsinstitute{Yerevan Physics Institute, Yerevan, Armenia}
{\tolerance=6000
V.~Chekhovsky, A.~Hayrapetyan, V.~Makarenko\cmsorcid{0000-0002-8406-8605}, A.~Tumasyan\cmsAuthorMark{1}\cmsorcid{0009-0000-0684-6742}
\par}
\cmsinstitute{Institut f\"{u}r Hochenergiephysik, Vienna, Austria}
{\tolerance=6000
W.~Adam\cmsorcid{0000-0001-9099-4341}, J.W.~Andrejkovic, L.~Benato\cmsorcid{0000-0001-5135-7489}, T.~Bergauer\cmsorcid{0000-0002-5786-0293}, S.~Chatterjee\cmsorcid{0000-0003-2660-0349}, K.~Damanakis\cmsorcid{0000-0001-5389-2872}, M.~Dragicevic\cmsorcid{0000-0003-1967-6783}, P.S.~Hussain\cmsorcid{0000-0002-4825-5278}, M.~Jeitler\cmsAuthorMark{2}\cmsorcid{0000-0002-5141-9560}, N.~Krammer\cmsorcid{0000-0002-0548-0985}, A.~Li\cmsorcid{0000-0002-4547-116X}, D.~Liko\cmsorcid{0000-0002-3380-473X}, I.~Mikulec\cmsorcid{0000-0003-0385-2746}, J.~Schieck\cmsAuthorMark{2}\cmsorcid{0000-0002-1058-8093}, R.~Sch\"{o}fbeck\cmsAuthorMark{2}\cmsorcid{0000-0002-2332-8784}, D.~Schwarz\cmsorcid{0000-0002-3821-7331}, M.~Sonawane\cmsorcid{0000-0003-0510-7010}, W.~Waltenberger\cmsorcid{0000-0002-6215-7228}, C.-E.~Wulz\cmsAuthorMark{2}\cmsorcid{0000-0001-9226-5812}
% ...
\par}
\vskip\cmsinstskip
\dag:~Deceased\\
$^{1}$Also at Yerevan State University, Yerevan, Armenia\\
$^{2}$Also at TU Wien, Vienna, Austria\\
% ...
\end{sloppypar}
\end{document}